\documentclass[sigconf]{acmart}
\copyrightyear{2019} 
\acmYear{2019} 
\setcopyright{acmcopyright}
\acmConference[ISCA '19]{The 46th Annual International Symposium on Computer Architecture}{June 22--26, 2019}{Phoenix, AZ, USA}
\acmBooktitle{The 46th Annual International Symposium on Computer Architecture (ISCA '19), June 22--26, 2019, Phoenix, AZ, USA}
\acmPrice{15.00}
\acmDOI{10.1145/3307650.3322247}
\acmISBN{978-1-4503-6669-4/19/06}

\newcommand{\ignore}[1]{}
\usepackage{amsmath}
\usepackage{algorithm}
\usepackage[noend]{algpseudocode}
\usepackage{tikz}
\usepackage{seqsplit}
\usepackage{subcaption}
\usepackage{microtype}

\newcommand*\circledB[1]{\tikz[baseline=(char.base)]{
		\node[shape=circle,fill=black,draw,inner sep=0.5pt] (char) {#1};}}

\begin{document}
%%%%%%%%%%%---SETME-----%%%%%%%%%%%%%
\title{OO-VR: NUMA Friendly \underline{O}bject-\underline{O}riented \underline{VR} Rendering Framework For Future NUMA-Based Multi-GPU Systems }

%
% The "author" command and its associated commands are used to define the authors and their affiliations.
% Of note is the shared affiliation of the first two authors, and the "authornote" and "authornotemark" commands
% used to denote shared contribution to the research.

\author{Chenhao Xie}
\email{chenhao.xie@pnnl.gov}
\affiliation{%
  \institution{Pacific Northwest National Lab (PNNL) and \\
                 University of Houston   }
}

\author{Xin Fu}
\email{xfu8@central.uh.edu}
\affiliation{%
  \institution{ECMOS Lab, ECE Department,\\
		University of Houston}
}

\author{Mingsong Chen}
\email{mschen@sei.ecnu.edu.cn}
\affiliation{%
  \institution{School of Computer Science and Software Engineering,\\
		East China Normal University}
}

\author{Shuaiwen Leon Song}
\email{Shuaiwen.Song@pnnl.gov}
\affiliation{%
 \institution{Pacific Northwest National Lab (PNNL) and \\
			  The University of Sydney}
}

\begin{abstract}
   With the strong computation capability, NUMA-based multi-GPU system is a promising candidate to provide sustainable and scalable performance for Virtual Reality (VR) applications and deliver the excellent user experience. However, the entire multi-GPU system is viewed as a single GPU under the single programming model which greatly ignores the data locality among VR rendering tasks during the workload distribution, leading to tremendous remote memory accesses among GPU models (GPMs). The limited inter-GPM link bandwidth (e.g., 64GB/s for NVlink) becomes the major obstacle when executing VR applications in the multi-GPU system.  
   By conducting comprehensive characterizations on different kinds of parallel rendering frameworks, we observe that distributing the rendering object along with its required data per GPM can reduce the inter-GPM memory accesses. However, this object-level rendering still faces two major challenges in NUMA-based multi-GPU system: (1) the large data locality between the left and right views of the same object and the data sharing among different objects and (2) the unbalanced workloads induced by the software-level distribution and composition mechanisms. 
   
   To tackle these challenges, we propose object-oriented VR rendering framework (OO-VR) that conducts the software and hardware co-optimization to provide a NUMA friendly solution for VR multi-view rendering in NUMA-based multi-GPU systems. We first propose an object-oriented VR programming model to exploit the data sharing between two views of the same object and group objects into batches based on their texture sharing levels. Then, we design an object aware runtime batch distribution engine and distributed hardware composition unit to achieve the balanced workloads among GPMs and further improve the performance of VR rendering. Finally, evaluations on our VR featured simulator show that OO-VR provides 1.58x overall performance improvement and 76\% inter-GPM memory traffic reduction over the state-of-the-art multi-GPU systems. In addition, OO-VR provides NUMA friendly performance scalability for the future larger multi-GPU scenarios with ever increasing asymmetric bandwidth between local and remote memory.  
\end{abstract}

\maketitle

\section{Introduction}

With the vast improvements in graphics technology, Virtual Reality (VR) is becoming a potential popular product for major high-tech companies such as Facebook \cite{OculusRift}, Google \cite{GoogleVR} and NVIDIA \cite{VRworks}. Different from normal PC or mobile graphics applications, VR promises a fully immersive experience to users by directly displaying images in front of users' eyes. Due to the dramatic experience revolution VR brings to users, 
the global VR market is expected to grow exponentially and generate \$30 billion annual revenue by 2022 \cite{VRmarket,VRmarket2}.  

Despite the growing market penetration, achieving true immersion for VR applications still faces severe performance challenges \cite{kanter2015graphics}. First, the display image must have a high pixel density as well as a broad field of views which requires a high display resolution. Meanwhile, the high-resolution image must be delivered at an extremely short latency so that users can preserve the continuous illusion of reality. However, the state-of-the-art graphics hardware -- the Graphics Processing Units (GPUs) in particular -- cannot meet these strict performance requirements \cite{kanter2015graphics}. %Significant performance improvements on GPUs are highly desired to satisfy numerous VR users.  
%To meet these requirements, significant performance improvement need to be made for graphics hardware -- the Graphics Processing Units (GPUs) in particular. 
Historically, GPUs gain performance improvements through integrating more transistors and scaling up the chip size, but these optimizations on single-GPU system can barely satisfy VR users due to the limited performance boost \cite{raffin2006pc}. Multi-GPU system with much stronger computation capability is a promising candidate to provide sustainable and scalable performance for VR applications \cite{raffin2006pc,Kim2017}. 

In recent years, the major GPU verdors combine multiple small GPU models (e.g., GPMs) to build a future multi-GPU system under a single programming model to provide scalable computing resources. They employ high speed inter-GPU links such as NVLINK \cite{VRSLi} and AMD Crossfire\cite{AMDcrossfire} to achieve fast data transmit among GPMs. The memory system and address mapping in this multi-GPU system are designed as a Non-Uniform Memory Access (NUMA) architecture to achieve 4x storage capacity over single-GPU system. 
The NUMA-based multi-GPU system employs shared memory space to avoid data duplication and synchronization overheads across the distributed memories \cite{arunkumar2017mcm}. In this study, we target the future multi-GPU system because it serves the VR applications more energy-efficiently than distributed multi-GPU system that employs separated memory space, and is becoming a good candidate for future mobile VR applications.  
%it achieves higher energy-efficiency than distributed multi-GPUs system that employs separated memory space, which plays a critical role in future mobile VR applications.  	
Since the entire system is viewed as a single GPU under the single programming model, the VR rendering workloads are sequentially launched and distributed to different GPMs without specific scheduling. Applying this naive single programming model greatly hurts the data locality among rendering workloads and incurs huge inter-GPM memory accesses, which significant constrain the performance of multi-GPU system for VR applications due to the bandwidth asymmetry between the local DRAM and the inter-GPM links. There have been many studies \cite{Kim2017,arunkumar2017mcm,milic2017beyond,Vinson2018remote} to improve the performance of NUMA-based multi-GPU system by minimizing the remote accesses. However, these solutions are still based on single programming model without considering the special data redundancy in VR rendering, hence, they cannot efficiently solve the performance bottleneck for VR applications.

Aiming to reduce the inter-GPM memory accesses, a straightforward method is employing parallel rendering frameworks \cite{OpenGLmultipipeSDK,humphreys2002chromium,eilemann2009equalizer,eilemann2018equalizer} to split the rendering tasks into multiple parallel sub-tasks under specific software policy before assigning to the multi-GPU system. Since these frameworks are originally designed for distributed multi-GPU system, a knowledge gap still exists on how to leverage parallel rendering programming model to efficiently execute VR applications in NUMA-based multi-GPU system. To bridge this gap, we first investigate three different parallel rendering frameworks (i.e. frame-level, tile-level and object-level). %when executing VR applications in NUMA-based multi-GPU system. 
By conducting comprehensive experiments on our VR featured simulator, we find that the object-level rendering framework that distributes the rendering object along with its required data per GPM can convert some remote accesses to local memory accesses. However, this object-level rendering still faces two major challenges in NUMA-based multi-GPU system: (1) a large number of inter-GPM memory accesses because it fails to capture the data locality between left and right view of the same object as well as the data sharing among different objects; (2) the serious workload unbalance among GPMs due to the inefficient software-level distribution and composition mechanisms.     

To overcome these challenges, we propose object-oriented VR rendering framework (OO-VR) that reduces the inter-GPM memory traffic by exploiting the data locality among objects. Our OOVR framework conducts the software and hardware co-optimizations to provide a NUMA friendly solution for VR multi-view rendering. First, we propose an object-oriented VR programming model that provides a simple software interface for VR applications to exploit the data sharing between the left and right views of the same object. The proposed programming model also automatically groups objects into batches based on their data sharing levels. Then, to combat the limitation of software-level solutions on workload distribution and composition, we design a object aware runtime batch distribution engine in hardware level to balance the rendering workloads among GPMs. We predict the execution time for each batch so that we can pre-allocate the required data of each batch to the local memory to hide long data copy latency. We further design the distributed composition unit in hardware level to fully utilize the rendering output units across all GPMs for best pixel throughput. %We evaluate the proposed design by executing real game benchmarks on our VR featured simulator. The results show that OOVR improves the overall performance by 1.58x on average and saves inter-GPM memory traffic by 76\% over the baseline case. In addition, we also provide two sensitivity studies to demonstrate that OOVR can potentially benefit the future larger multi-GPU scenario with ever increasing asymmetric bandwidth between local and remote memory. 
To summarize, the paper makes following contributions:
\begin{itemize}
	\item We investigate the performance of future NUMA-based multi-GPU systems for VR applications, and find that the inter-GPM memory accesses are the major performance bottleneck.  
	\item We conduct comprehensive characterizations on major parallel rendering frameworks, and observe that the data locality among rendering objects can help to significantly reduce the inter-GPM memory accesses but the state-of-the-art frameworks and multi-GPU systems fail to capture this interesting feature. 	
	\item We propose a software and hardware co-designed \textit{O}bject-\textit{O}riented \textit{VR} (OO-VR) rendering framework that leverages the data locality feature to convert the remote inter-GPM memory accesses to local memory accesses.
	\item We further build a VR featured simulator to evaluate our proposed design by rendering VR enabled real-world games with different resolutions. The results show that OO-VR achieves 1.58x performance improvement and 76\% inter-GPM memory traffic reduction over the state-of-the-art multi-GPU system. With its nature of NUMA friendly, OO-VR exhibits strong performance scalability and potentially benefits the future larger multi-GPU scenarios with ever increasing asymmetric bandwidth between local and remote memory. 
	
\end{itemize} 
\section{Background and Motivation}
\label{sec:Background}

\subsection{Multi-View Rendering in Virtual Reality}

\begin{figure}[t]
	\begin{center}
		\includegraphics[width=0.45\textwidth]{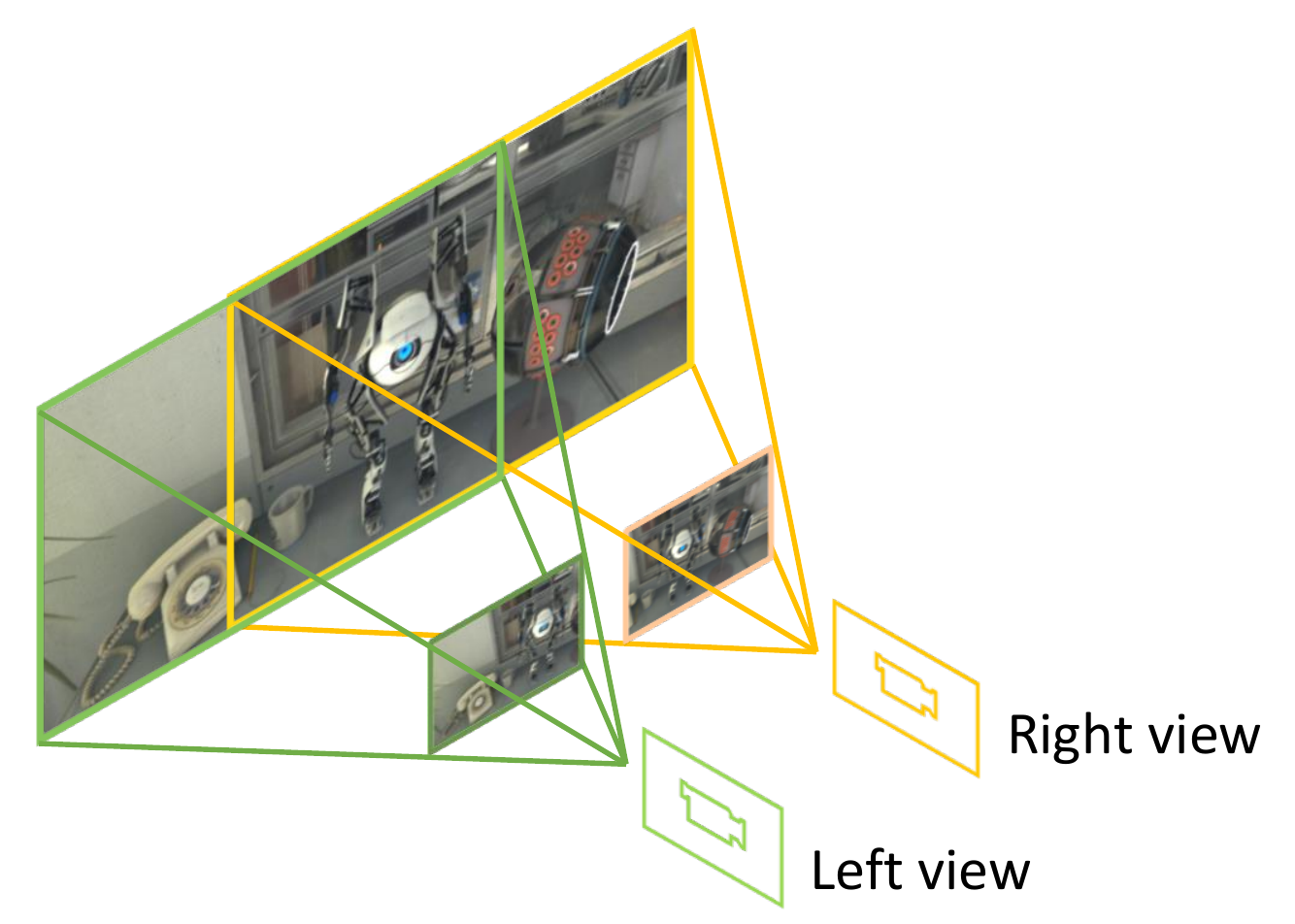}
			\vspace{-0.3cm}
		\caption{Rendering the VR world into left and right views. (Frame is captured from The LAB\cite{theLab})}
		
		\label{fig:sharedobject}
	\end{center}
	\vspace{-0.5cm}
\end{figure} 

\begin{figure*}[t]
	\begin{center}
		\includegraphics[width=\textwidth]{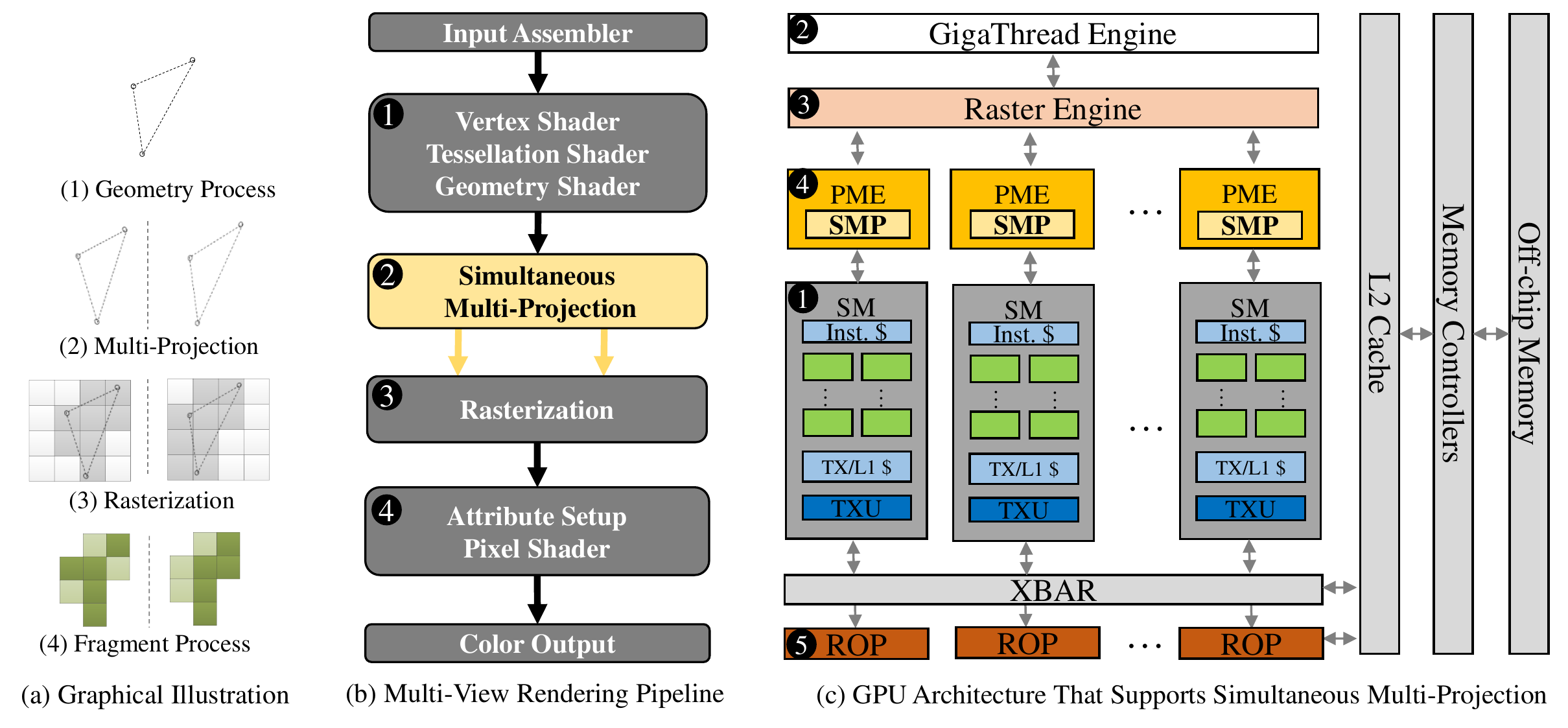}
			\vspace{-0.8cm}
		\caption{Multi-view VR rendering process with simultaneous multi-projection (SMP) enabled: (a) the output of each rendering step; (b) The overview of the 4-step rendering process; and (c) GPU architecture for SMP-enabled VR rendering. Modern GPUs employ unified shader model that processes all shaders in programmable Streaming Multiprocessors (SMs). They also feature a new fixed function unit inside each Polymorph Engine (PME) to support SMP. }
		
		\label{fig:singlepass}
	\end{center}
	\vspace{-0.3cm}
\end{figure*} 

In contrast to other traditional graphics applications, the state-of-the-art VR applications employ \textit{Head-Mounted-Display} (HMD), or VR helmet, to directly present visuals to users' eyes. To display 3D objects in VR, a pair of frames  (i.e., \textit{stereoscopic frames}) are generated for both left and right eyes by projecting the scene onto two 2D plate images. This process is referred as stereo rendering in computer graphics. Figure \ref{fig:sharedobject} shows an example of such VR projection. The green and yellow boxes represent the rendering process for left and right views, respectively, creating two display images for the HMD. Stereo rendering requires two concurrent rendering process for the two eyes' views, resulting in doubled amount of workload for the VR pipeline. Due to the observation that some objects in the scene (e.g. the robot in Figure \ref{fig:sharedobject}) are shared by two eyes, mainstream graphics engines such as NVIDIA and UNITY employ \textit{simultaneous multi-projection} (SMP) to generate the left and right frames simultaneously through single rendering process \cite{GTX1080,Unity,chen2010real,de2010gpu}. This significantly reduces workload redundancy and achieves substantial performance gain.

%\subsection{VR Rendering Pipeline}
Based on the conventional three-step rendering process (i.e., Geometry Process, Rasterization and Fragment Process) defined by modern graphics application programming interface (API) \cite{openGL, D3D}, VR rendering inserts multi-projection process after the geometry process and prior to the Rasterization, shown in Figure \ref{fig:singlepass}(a). Thus, when SMP is enabled, VR rendering process is composed of four steps, detailed in Figure \ref{fig:singlepass}(b). Basically, VR rendering begins from reading the application-issued vertex from GPU memory. During the geometry process \circledB{\color{white}1}, the vertex shader calculates the 3D coordinates of the vertex and assembles them into primitives (i.e. triangles in Figure \ref{fig:singlepass}(a)-(1)). After that, the generated triangles pass through the geometry-related shaders which perform clipping, face culling and tessellation to generate extra triangles and remove non-visible triangles. Then, the SMP step \circledB{\color{white}2} is responsible for generating multiple projections of a single geometry stream. In other words, GPU executes geometry process only once but produces two positions for each triangle (Figure \ref{fig:singlepass}(a)-(2)).
These triangles are then streamed into the rasterization stage \circledB{\color{white}3} to generate fragments (Figure \ref{fig:singlepass}(a)-(3)), each of which is equivalent to a pixel in a 2D image. Finally, the fragment process \circledB{\color{white}4} generates pixels by calculating the corresponding fragment attributes to determine their colors and texture (Figure \ref{fig:singlepass}(a)-(4)). The output pixels will be written into the frame buffer in GPU memory for displaying.

\subsection{SMP Featured GPU Architectures} 
Traditionally, GPUs are designed as the special-purpose graphics processors for performing modern rendering tasks. Figure \ref{fig:singlepass}(c) shows a SMP supported GPU architecture which models the recent NVIDIA Pascal GPUs\cite{GTX1080}. 
It consists of several programmable streaming multiprocessors (SMs) \circledB{\color{white}1}, some fixed function units such as the GigaThread Engine \circledB{\color{white}2}, Raster Engine \circledB{\color{white}3}, Polymorph Engine (PME) \circledB{\color{white}4}, and Render Output Units (ROPs) \circledB{\color{white}5}. Each SM \circledB{\color{white}1} is composed of a unified texture/L1 cache (TX/L1 \$), several texture units (TXU) and hundreds of shader cores that execute a variety of graphics shaders (e.g., the functions in both geometry and fragment process). The GigaThread Engine \circledB{\color{white}2} distributes the rendering workloads among PMEs if there are adequate computing resources. The raster engine \circledB{\color{white}3} is a hardware accelerator for rasterization process. Each PME \circledB{\color{white}3} conducts input assembler, vertex fetching, and attribute setup. To support multi-view rendering, NVIDIA Pascal architecture integrates an SMP engine into each PME. The SMP engine is capable of processing geometry for two different viewports which are the projection centers for the left and right views. In other words, it duplicates the geometry process from left to right views through changing the projection centers instead of executing the geometry process twice. Finally, the Render Output Units (ROPs) \circledB{\color{white}5} perform anti-aliasing, pixel compression and color output. As Figure \ref{fig:singlepass}(c) illustrates, all the SMs and ROPs share a L2 cache and read/write data to the off-chip memory through the memory controllers, each of which is paired to one memory channel. Prior to rendering, GPU memory contents such as framebuffer and texture data are pre-allocated in GPU's off-chip memory. Programmers can manually manage the memory allocation using the graphics APIs such as OpenGL and Direct3D\cite{openGL, D3D}.

%When a graphics application issues a draw command, the GigaThread Engine distributed the rendering workloads to each PME for input assembler and vertex fetching if there are plenty of computing resources. Then, the programing shaders in geometry process will be executed in each SM to generate triangles. These triangles are then sent back to PME and raster engine for SMP and rasterization, respectively. After that, these triangles are duplicated and translated into fragments. Finally, these fragments are distributed into SMs to execute pixel shade and the output colors are wrote back to GPU off-chip memory though ROPs according to their screen space coordinates. 

%\subsection{GPU Performance Scaling}
Although recent generations of GPUs have shown capability to deliver good gaming experiences and also gradually evolved to support SMP, it is still difficult for them to satisfy the extremely high demands on rendering throughput from immersive VR applications. The human vision system has both wide field of view (FoV) and incredibly high resolution when perceiving the surrounding world; the requirement for enabling an immersive VR experience is much more stringent than that for PC gaming. Table \ref{table:PCvsVR} lists the major differences between PC gaming and stereo VR \cite{kanter2015graphics}. As it demonstrates, stereo VR requires GPU to deliver 116 (58.32$\times$2) Mpixels within 5 ms. Missing the rendering deadline will cause frame drop which significantly damages VR quality. Although the VR vendors today employ frame re-projection technologies such as Asynchronous Time Warp (ATW)\cite{van2016asynchronous, TimewarpLatency} to artificially fill in dropped frames, they cannot fundamentally solve the problem of rendering deadline missing due to little consideration on users' perception and interaction. Thus, improving the overall rendering efficiency is still the highest design priority for modern VR-oriented GPUs \cite{oculusATW}.

\begin{scriptsize}
	\begin{table}[t] \footnotesize
	    	    \captionsetup{labelfont = bf}
		\centering
		\caption{\bf Differences Between PC Gaming and VR}
		\vspace{-0.3cm}
		\label{table:PCvsVR}
		\begin{tabular}{|l|l|l|}
			\hline
			& {\bf Gaming PC} & {\bf Stereo VR} \\
			\hline
			Display & 2D LCD panel & Stereo HMD\\
			\hline
			Field of View (FoV) & 24-30" diagonal & 120$^\circ$ horizontally\\
			& & 135$^\circ$ vertically \\
			\hline
			Number of Pixel & 2-4 Mpixels & 58.32x2 Mpixels\\ 
			\hline
			Frame latency & 16-33 ms & 5-10 ms\\
			\hline
		\end{tabular}	
		\vspace{-0.2cm}	
	\end{table}	
\end{scriptsize}  

%Historically, GPU performance improvements have been tightly coupled to transistor scaling. However, as Moore?s law slows down recently, the number of transistors per die no longer grows at historical rates, and it takes longer to upgrade the performance of a single-chip GPUs. For instance, the NVIDIA GTX 980 (Maxwell) which was released in 2015 had double the number of SMs comparing to the GTX 680 (Kepler) which was released two years before it \cite{GTX980} while GTX 1080Ti (Pascal) which was released in 2017 only increases 50\% SMs than GTX 980 \cite{GTX1080}. 
%In order to satisfy the requirement for immersive VR applications, one straightforward solution to improve GPU performance is employ multiple GPU architecture. 

\begin{figure}[t]
	\begin{center}
		\includegraphics[width=0.48\textwidth]{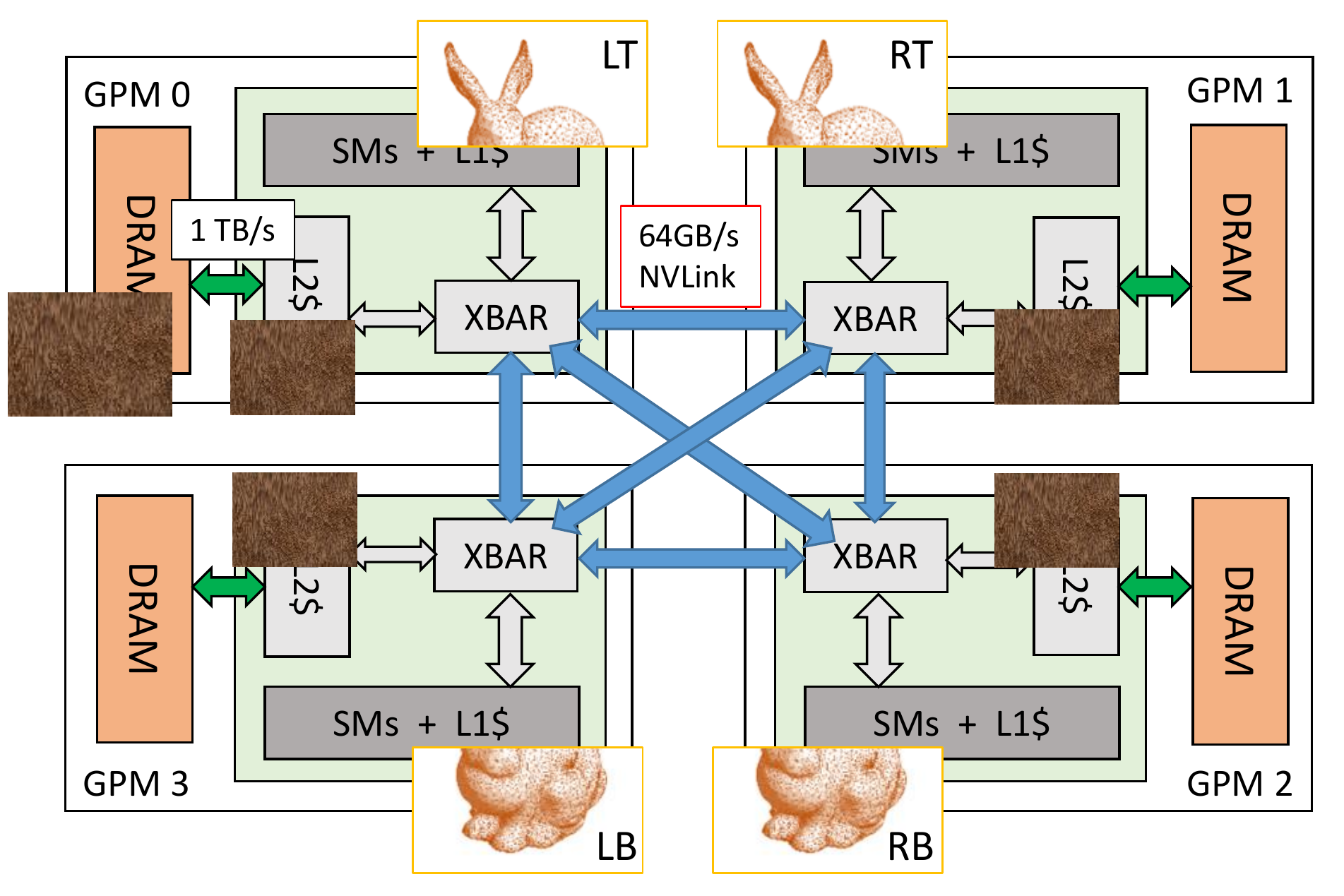}
		\vspace{-0.7cm}
		\caption{The Overview of the multi-GPU architecture. Distributed rendering tasks for the same object causes significant remote memory access and data duplication. }
		
		\label{fig:MCM}
	\end{center}
	\vspace{-0.4cm}
\end{figure} 

\subsection{NUMA-Based Multi-GPU System and Its Performance Bottleneck}
In recent years, major GPU vendors such as NVIDIA have proposed to integrate multiple easy-to-manufacture GPU chips at package level (i.e., multi-chip design)\cite{arunkumar2017mcm} or at system level\cite{milic2017beyond,Vinson2018remote} using high bandwidth interconnection technologies such as Grand-Reference Signaling (GRS)\cite{poulton20130} or NVLinks\cite{VRSLi}, in order to address future chip density saturation issue.
Figure \ref{fig:MCM} shows the overview of the multi-GPU architecture which consists of four GPU models (i.e., GPMs). In terms of compute capability, each GPM is configured to resemble the latest NVIDIA GPU architecture (e.g., Figure \ref{fig:singlepass}(c)). Inside each GPM, SMs are connected to the GPM local memory hierarchy including a local memory-side L2 cache and off-chip DRAM, via an XBAR. In the overall multi-chip design (MCM-GPU), XBARs are interconnected through high speed links such as NVLinks to support the communication among different GPMs. This multi-GPU system generally acts as a large single GPU; its memory system and address mapping are designed as a Non-Uniform Memory Access (NUMA) architecture. This design also reduces the programming complexity (e.g., unified programming model similar to CUDA) for GPU developers.  

\textbf{Future Multi-GPU System Bottleneck for VR Workloads.} As previous works \cite{arunkumar2017mcm,milic2017beyond} have indicated, data movement among GPMs will become the major obstacle for the continued performance scaling in these future NUMA-based multi-GPU systems. This situation is further exacerbated when executing VR applications caused by the large data sharing among GPMs. Due to the nature of view redundancy in VR applications, the left and right views may include the same object (e.g., the rabbit in Figure \ref{fig:MCM}) which require the same texture data. However, to effectively utilize the computing resources from all the GPMs in such multi-GPM platforms, the rendering tasks for left and right views will be distributed to different groups or islands of GPMs in a more balanced fashion; each view will then be further broken into smaller pieces and distributed to the individual GPMs of that group. This naive strategy could greatly hurt data locality in the SMP model. For example, if the basic texture data used to describe the rabbit in Figure \ref{fig:MCM} is stored in the local memory of GPM\_0, other GPMs need to issue remote memory accesses to acquire this data. Due to the asymmetrical bandwidth between the local DRAM (e.g., 1TB/s) and inter-GPM NVLink (e.g., 64GB/s), the remote memory access will likely become one of the major performance bottlenecks in such multi-GPU system design. More sophisticated parallel rendering frameworks such as OpenGL Multipipe SDK \cite{OpenGLmultipipeSDK}, Chromium \cite{humphreys2002chromium} and Equalizer\cite{eilemann2009equalizer,eilemann2018equalizer}, are designed for distributed environment where they separate memory space and the memory data need to be duplicated in each memory which greatly limits the storage capacity on our NUMA-based multi-GPU systems. Thus, employing them on our architecture requires further investigation and characterization. We will show this study in Section \ref{sec:Character}.  

\begin{figure}[t]
	\begin{center}
		\includegraphics[width=0.48\textwidth]{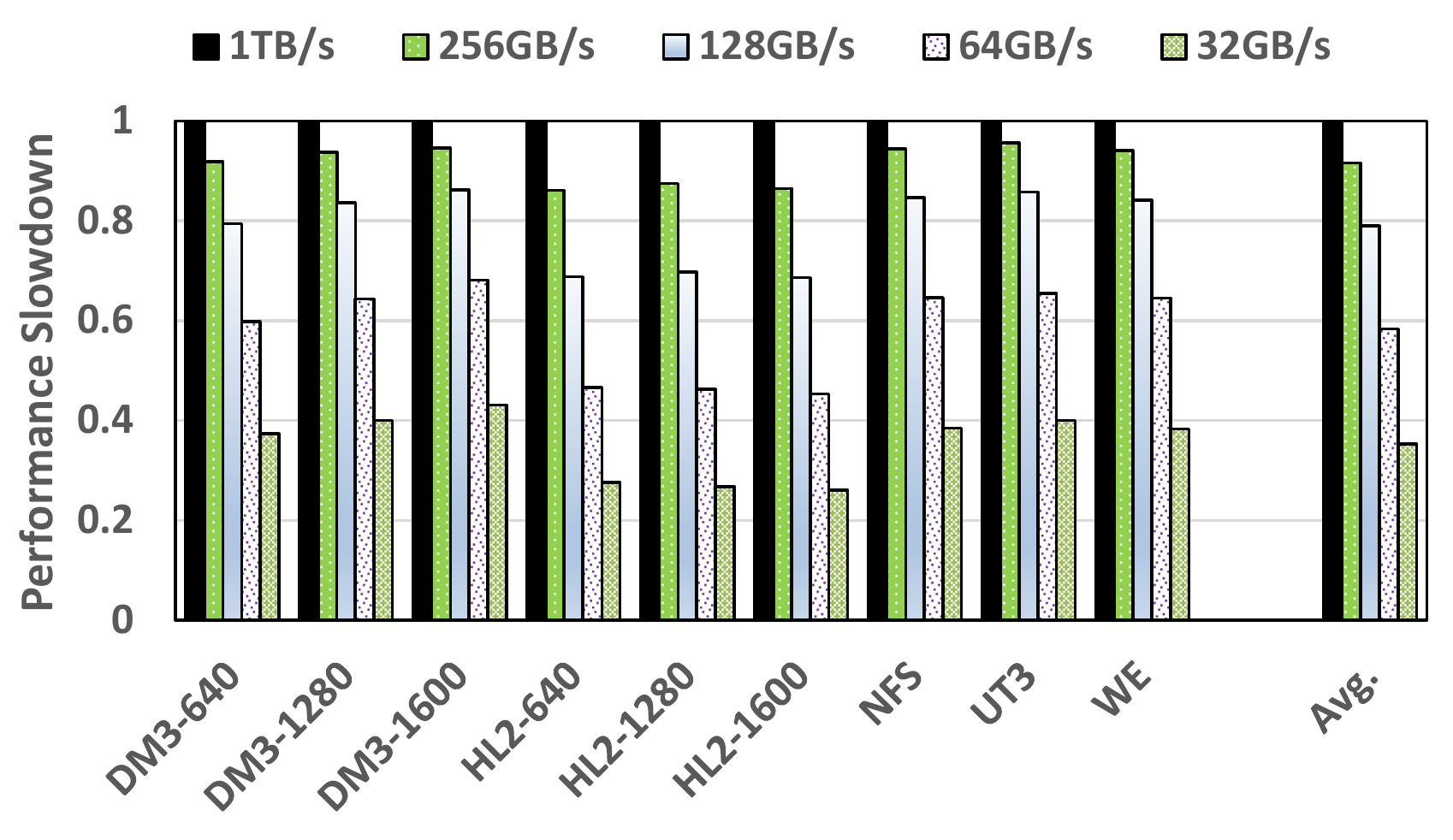}
		\vspace{-0.7cm}
		\caption{Normalized performance sensitivity to inter-GPM link bandwidth for a 4-GPM Multi-GPU system.}
		
		\label{fig:SL}
	\end{center}
	\vspace{-0.3cm}
\end{figure}

Figure \ref{fig:SL} presents the performance of a 4-GPM multi-GPU system as the bandwidth of inter-GPM links is decreased from 1TB/s to 32GB/s (refer to Section \ref{sec:Methodology} for experimental methodology). We can observe that the rendering performance is significantly limited by the bandwidth. On average, applying 128GB/s, 64GB/s and 32GB/s inter-GPM bandwidth results in 22\%, 42\% and 65\% performance degradation compared to the baseline 1TB/s bandwidth, respectively. Although improving the inter-GPM bandwidth is a straightforward method to tackle the problem, it has proven difficult to achieve due to additional silicon cost and power overhead \cite{arunkumar2017mcm}.  %\textcolor{red}{[Do you have a more straightforwared way to show the results? E.g., what is the percentage of time on inter-GPM communication for 32GB/s inter-GPM link.]} 

This motivates us to provide software-hardware co-design strategies to enable "true" immersive VR experience for future users via significantly reducing the inter-GPM traffic and alleviating the performance bottleneck of executing VR workloads on future multi-GPU platforms. We believe this is the first attempt to co-design at system architecture level for eventually realizing future \textit{planet-scale} VR.

%We also observe the limitation of inter-GPM bandwidth causes similar performance degradations for different benchmarks and resolution. That means reducing the NUMA bottleneck has the highest priority to improve the performance on Multi-GPU system, which is the major object in this work.
      
%Recent works for multi-GPU architecture proposed to distribute rendering tasks as closed as possible to its data \cite{Kim2017}. However, the modern VR application usually employ a single large page to store the texture data which stored in one DRAM, distribute all tasks within the same GPM will cause low utilization in other GPMs. 
%Previous works also suggest to replicate the remote data in the local memory system (i.e. L2 cache and memory) of each GPM to reduce remote GPU memory access \cite{milic2017beyond}. However, these technologies increase the capability pressure for GPU memory. In VR application, the reducing GPU memory capacity directly limit the rendering resolution, the number of perceived object and texture detail. Increasing the capability pressure means sacrifice the immersive quality which is unworthy in VR applications. 

\section{Experimental Methodology}
\label{sec:Methodology}
We investigate the performance impact of multi-GPU system for virtual reality by extending ATTILA-sim \cite{attila}, a cycle-level rasterization-based GPU simulator which covers a wide spectrum of graphics features on modern GPUs. The model of ATTILA-sim is designed upon \emph{boxes} (a module of functional pipeline) and \emph{signals} (simulating the interconnect of different components). Because the current ATTILA-sim models an AMD TeraScale2 architecture \cite{houston2008anatomy}, it is difficult to configure it using the same amount of SMs as NVIDIA Pascal-like architectures \cite{GTX1080}. To fairly evaluate the design impact, we accordingly scale down other parameters such as the number of ROPs and L2 cache. Similar strategies have been used to study modern graphics architectures in previous works\cite{PIMrendering,Perception3D,PIMVR}. The GPM memory system consists of two level cache hierarchy and a local DRAM. The L1 cache is private to each SM while the L2 cache are shared by all the SMs. Table \ref{table:configuration} shows the simulation parameters applied in our baseline multi-GPU system. 

\begin{scriptsize}
	\begin{table}[t]\footnotesize
	    \captionsetup{labelfont = bf}
		\centering
		\caption{\bf BASELINE CONFIGURATION}
		\vspace{-0.3cm}
		\label{table:configuration}
		\centering
		\begin{tabular}{|l|l|}
			\hline
			GPU frequency & 1GHz\\
			\hline
			Number of GPMs & 4 \\
			\hline
			Number of SMs & 32, 8 per GPMs \\
			\hline
			SM configuration & 64 shader core per SM \\
			& 128KB Unified L1 cache \\
			& 4 texture unit\\
			\hline
			Texture filtering configuration &  16x Anisotropic filtering\\
			\hline
            Raster Engine & 16x16 tiled rasterization\\
            \hline
            Number of ROP & 32, 8 per GPMs \\
            \hline		
			L2 cache & 4MB in total, 16-ways\\
			\hline
			Inter-GPU interconnect & 64GB/s NVLink\\
			&uni-directional\\
			\hline
			Local DRAM bandwidth & 1TB/s \\
			\hline
		\end{tabular}
	\end{table}
\end{scriptsize}

\begin{figure}[t]
	\begin{center}
		\includegraphics[width=0.43\textwidth]{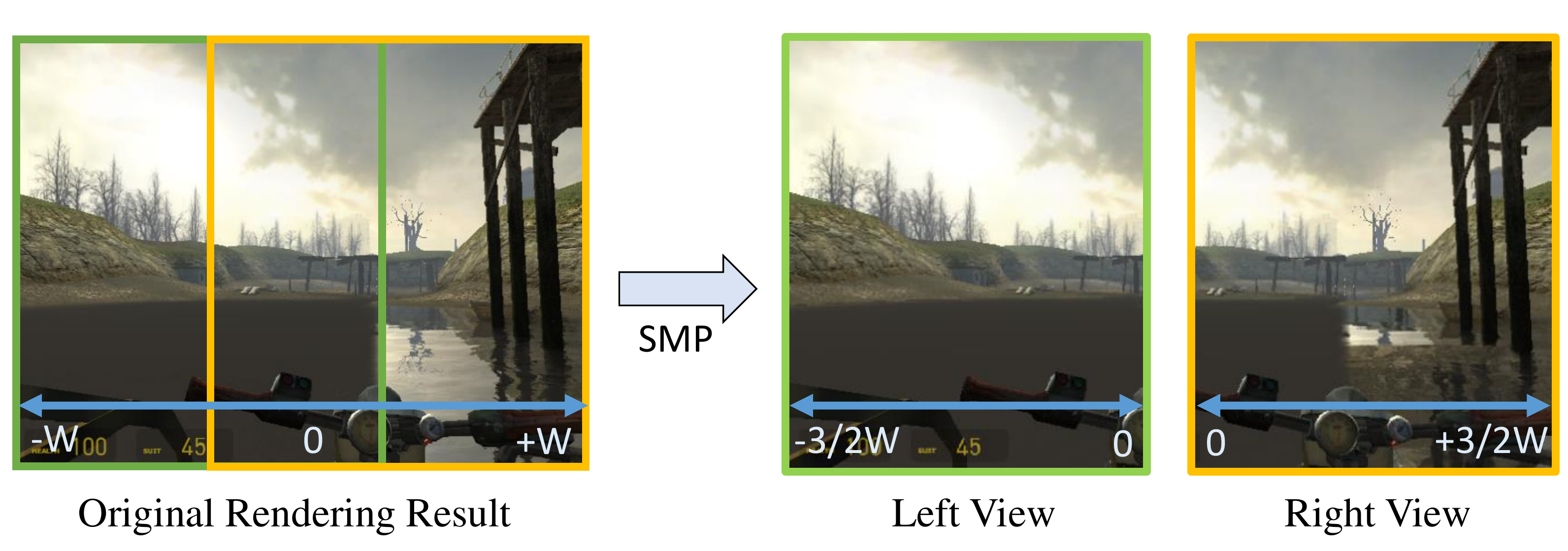}
			\vspace{-0.3cm}
		\caption{The original rendering frame (left) and results with SMP enabled (right).}
		\label{fig:Implement}
	\end{center}
	\vspace{-0.5cm}
\end{figure}

In order to support multi-view VR rendering, we implement the SMP engine in ATTILA-sim based on the state-of-the-art SMP technology \cite{de2010gpu,chen2010real} which re-projects the triangles in left view to right using updated viewport. Figure \ref{fig:Implement} shows the rendering example of Half-Life 2 after enabling SMP in ATTILA-sim. Our SMP engine first gathers the X coordinate of the display frame which is from -W to +W, where W is a coordinate offset parameter. Then, it duplicates each triangle generated from the geometry process. After that, the SMP engine shifts the viewport of the rendering object by half of W, left or right depending on the eye. The SMP engine can also re-project the triangle based on user-defined viewports for left and right views. Finally, we modify the triangle clipping to prevent the spill over into the opposite eye. We validated the implementation of the SMP engine in ATTILA-sim by comparing the triangle number, fragment number and performance improvement with that from executing VR benchmarks on the state-of-the-art GPUs (e.g. Sponza and San Mangle in NVIDIA VRWork \cite{VRworks}) on NVIDIA GTX 1080 Ti). Specifically, we observe that the added SMP rendering on ATTILA-sim can provide a 27\% speed up over the sequential rendering on two views. 

We also model the inter-GPU interconnect as high bandwidth point-to-point NVLinks with 64GB/s bandwidth (one direction). We assume each GPM has 6 ports and each pair of ports is used to connect two GPMs, indicating that the intercommunication between two GPMs will not be interfered by other GPMs. Based on the sensitivity study shown in Figure \ref{fig:SL}, we configure the inter-GPM link bandwidth as 64GB/s bandwidth. Following the common GPU design, each ROP in our simulation outputs 4 pixels per cycle to the framebuffer. To further alleviate the remote memory access latency on the NUMA-based baseline architecture, we employ the state-of-the-art First-Tough (FT) page placement policy and remote cache scheme \cite{arunkumar2017mcm} to create a fair baseline evaluation environment.  

%We evaluate a set of benchmarks from the graphics API traces from ATTILA, covering different rendering libraries and 3D engines.

\begin{scriptsize}
	\begin{table}[t]\footnotesize
	    \captionsetup{labelfont = bf}
		\centering
		\caption{\bf BENCHMARKS}
		\vspace{-0.3cm}
		\begin{tabular}{|l|l|l|l|l|p{0.48\textwidth}|}
			\hline
			\bf Abbr. &\bf Names & \bf Library & \bf Resolution & \bf \#Draw  \\
			\hline
			\hline
			DM3 & Doom 3 & OpenGL\cite{openGL} &1600x1200 & 191 \\
			& & &1280x1024 &  \\
			& & &640x480 &  \\
			\hline
			HL2 & Half-Life 2 & DirectX\cite{D3D} & 1600x1200 & 328 \\
			& & &1280x1024 &  \\
			& & &640x480 &  \\
			\hline
		    NFS & Need For & DirectX & 1280x1024 &  1267 \\
		    & Speed & & &\\
			\hline
			UT3 & Unreal & DirectX & 1280x1024 &  876 \\
			& Tournament 3 & & &\\
			\hline
			WE & Wolfenstein &DirectX &640x480 & 1697 \\	
			\hline
		\end{tabular}
		\label{table:benchmarks}
	\end{table}
\end{scriptsize}

\begin{figure*}[t!]
	\centering
	\begin{minipage}[h]{0.245\textwidth}
		\includegraphics[width=1\textwidth]{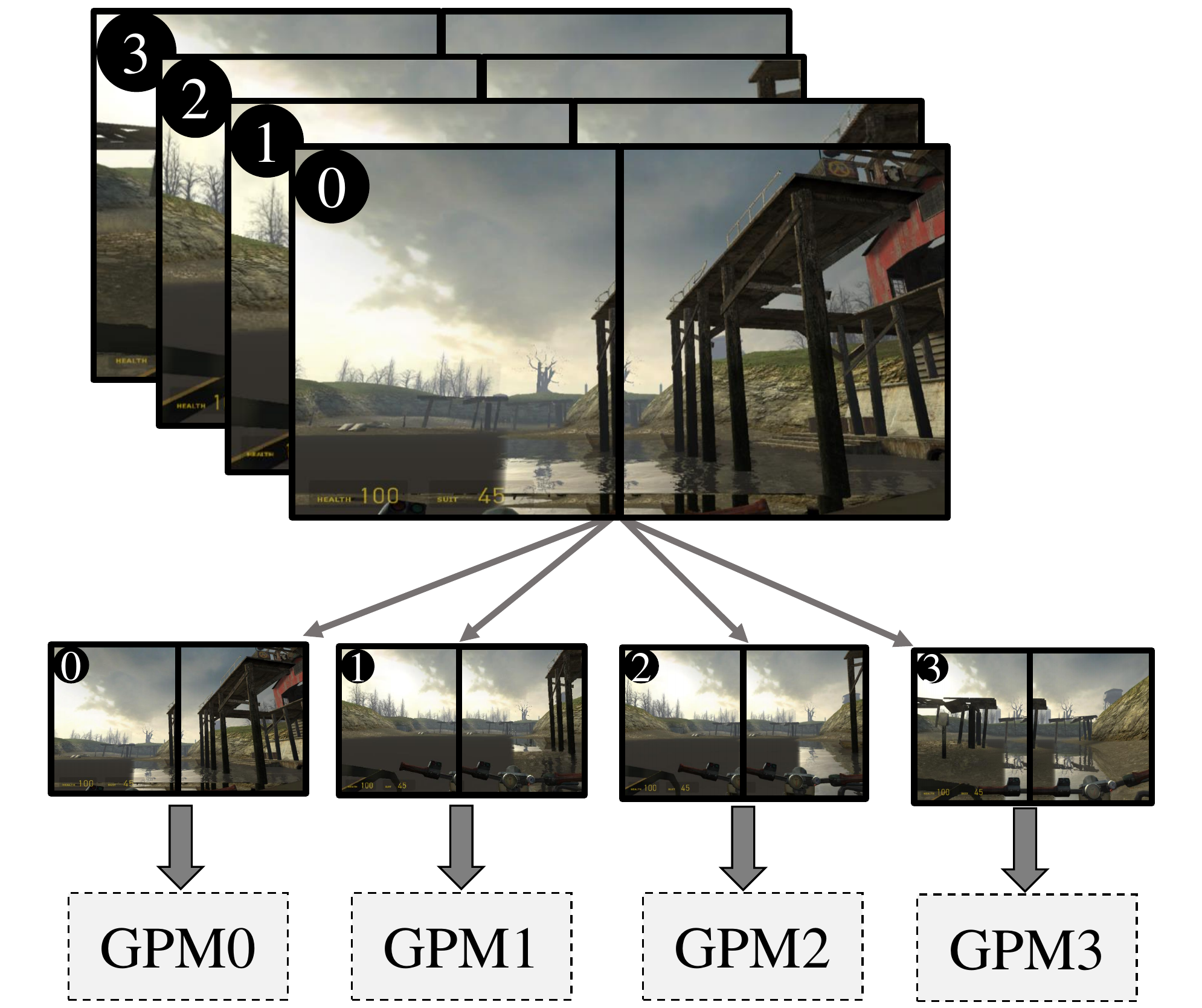}
		\vspace{-0.4cm}\subcaption{AFR}\label{fig:AFRimplement}		
	\end{minipage}
	\begin{minipage}[h]{0.235\textwidth}
		\includegraphics[width=1\textwidth]{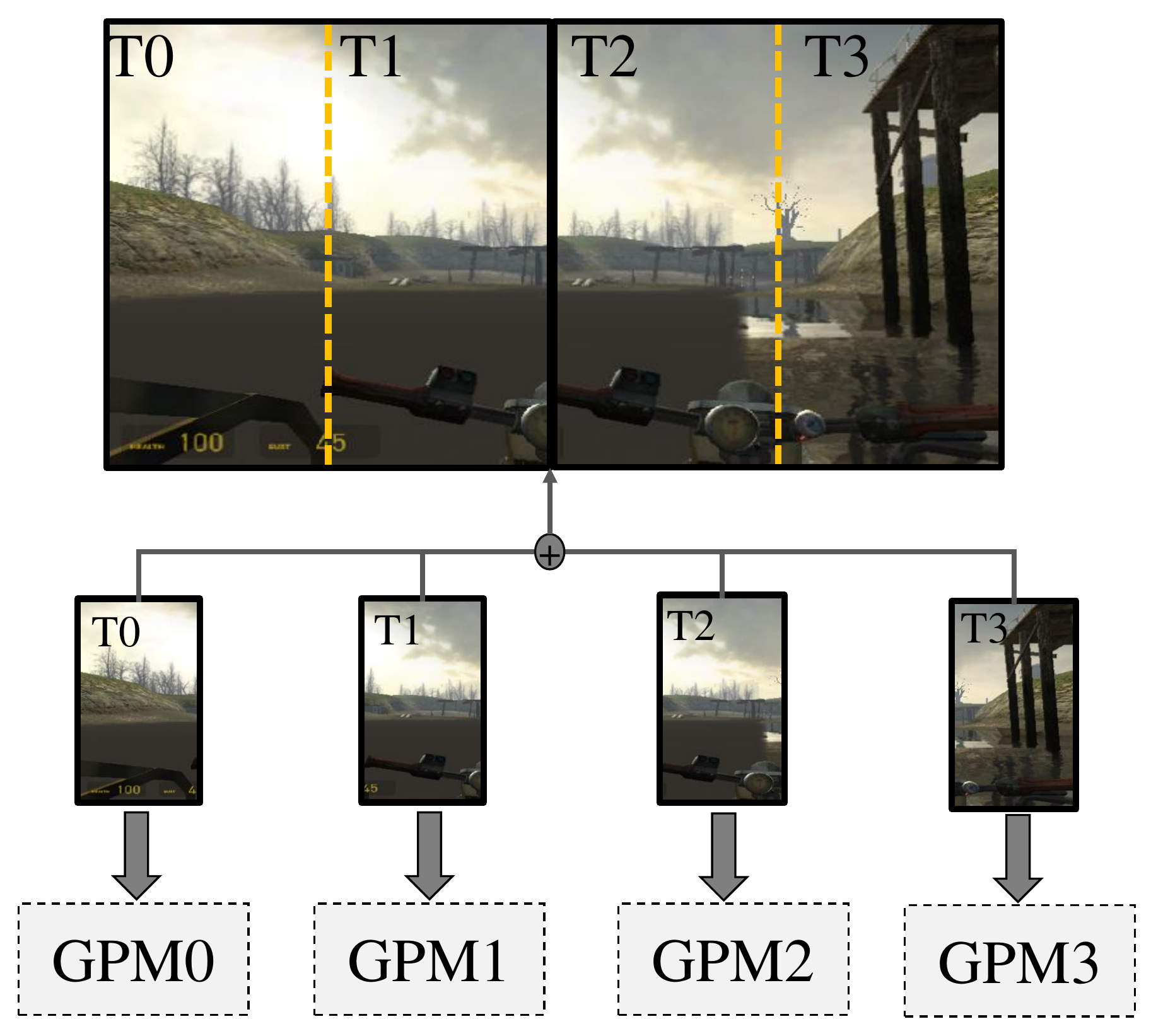}
		\vspace{-0.45cm}\subcaption{Tile-level SFR (V)}\label{fig:TBR1}		
	\end{minipage}
	\begin{minipage}[h]{0.24\textwidth}
		\includegraphics[width=1\textwidth]{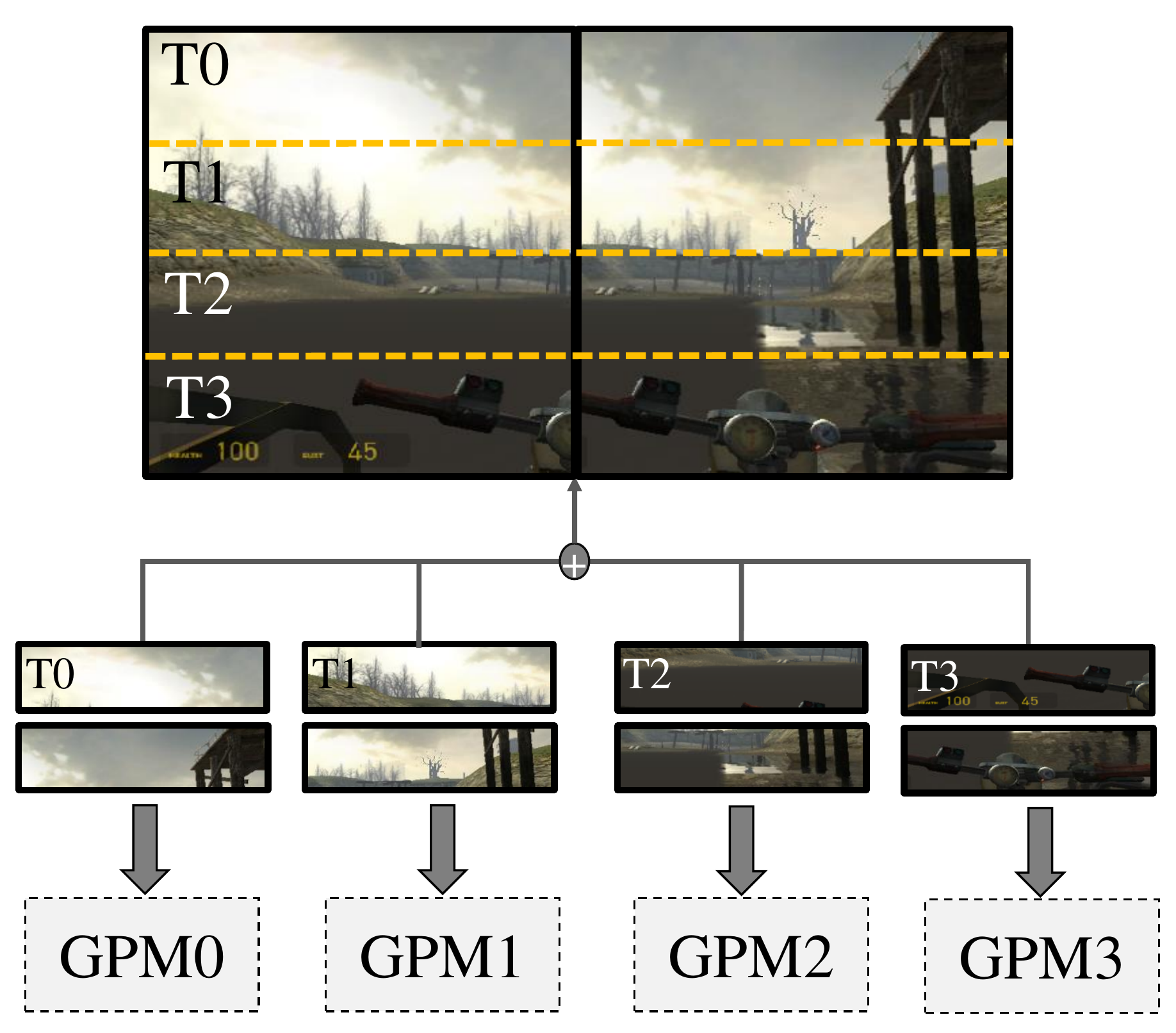}
		\vspace{-0.45cm}\subcaption{Tile-level SFR (H)}\label{fig:TBR2}
	\end{minipage}
	\begin{minipage}[h]{0.235\textwidth}
		\includegraphics[width=1\textwidth]{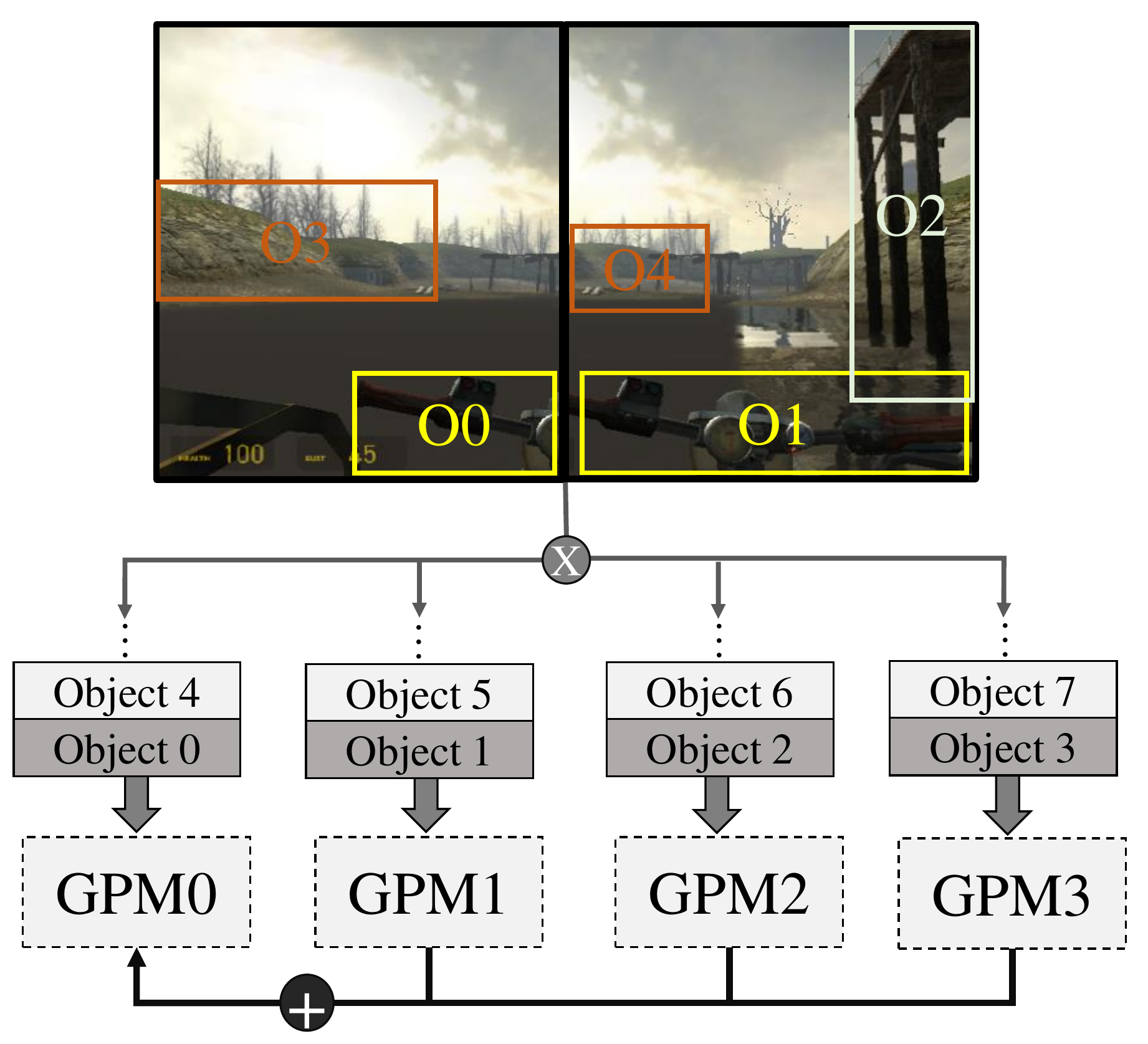}
		\vspace{-0.55cm}\subcaption{Object-level SFR}\label{fig:OBR}
	\end{minipage}
	\vspace{-0.3cm}
	\caption{ Three types of parallel rendering schemes for parallel VR applied on future NUMA-based Multi-GPU systems.}
	\vspace{-0.2cm}
\end{figure*}

Table \ref{table:benchmarks} lists the set of graphics benchmarks employed to evaluate our design. This set includes five well-known 3D games, covering different rendering libraries and 3D engines. We also list the original rendering resolution and the number of draw commands for these benchmarks. Two benchmarks (\emph{Doom3 and Half-Life 2}) from the table are rendered with a range of resolutions (1600$\times$1200, 1280$\times$1024, 640$\times$480) while for other games we adopt 1280x1024 resolution if it is available and supported by the simulator. In order to feature these PC games as VR applications, we modify the ATTILA Common Driver Layer (ACDL) to enable the multi-view rendering. In our experiments, we let all the workloads run to completion to generate the accurate frames on the simulator and gather the average frame latency for each game.

 \begin{figure}[t]
 	\centering
 	\begin{minipage}[h]{0.225\textwidth}
 		\includegraphics[width=1\textwidth]{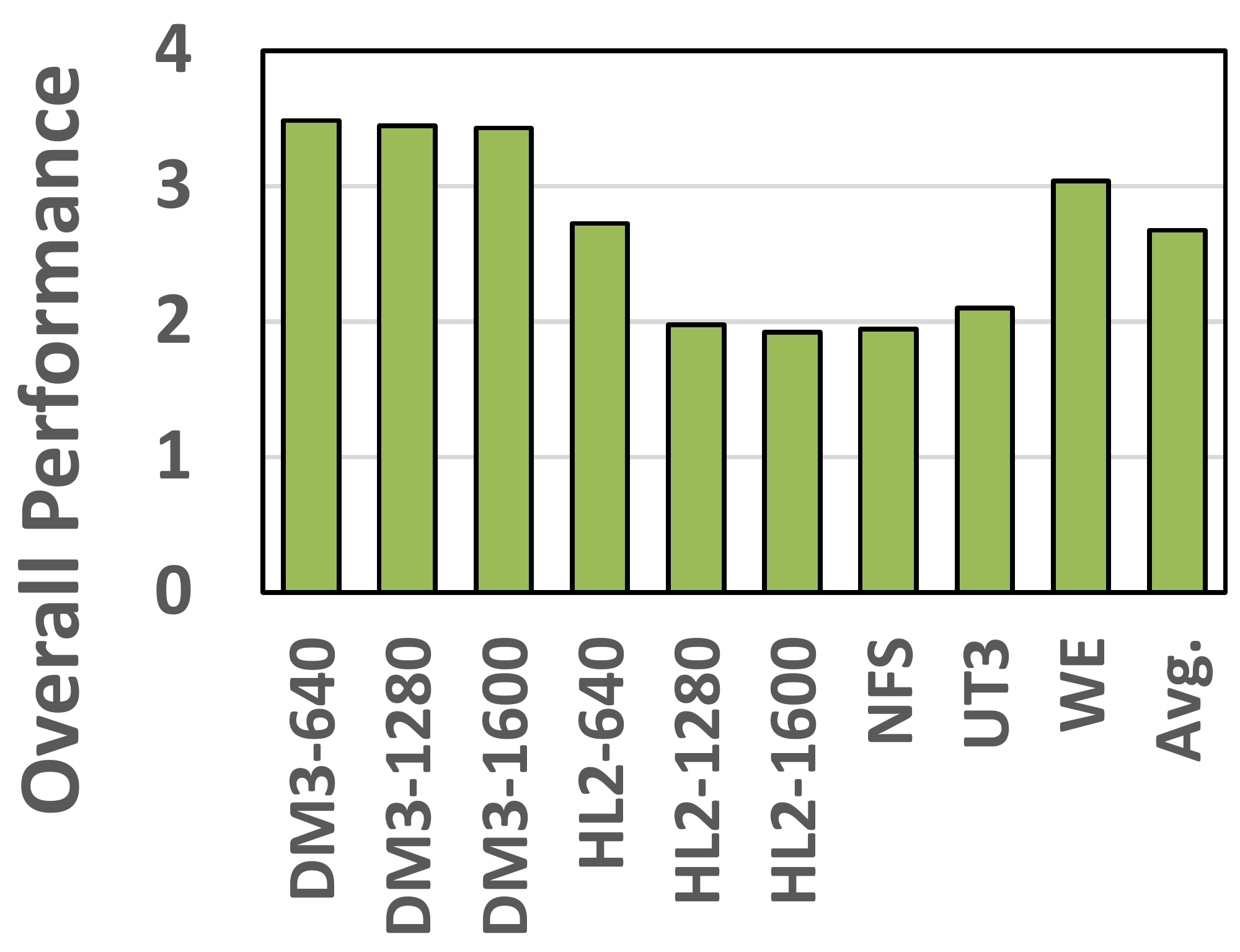}
 		\vspace{-0.5cm}		
 	\end{minipage}
 	\begin{minipage}[h]{0.24\textwidth}
 		\includegraphics[width=1\textwidth]{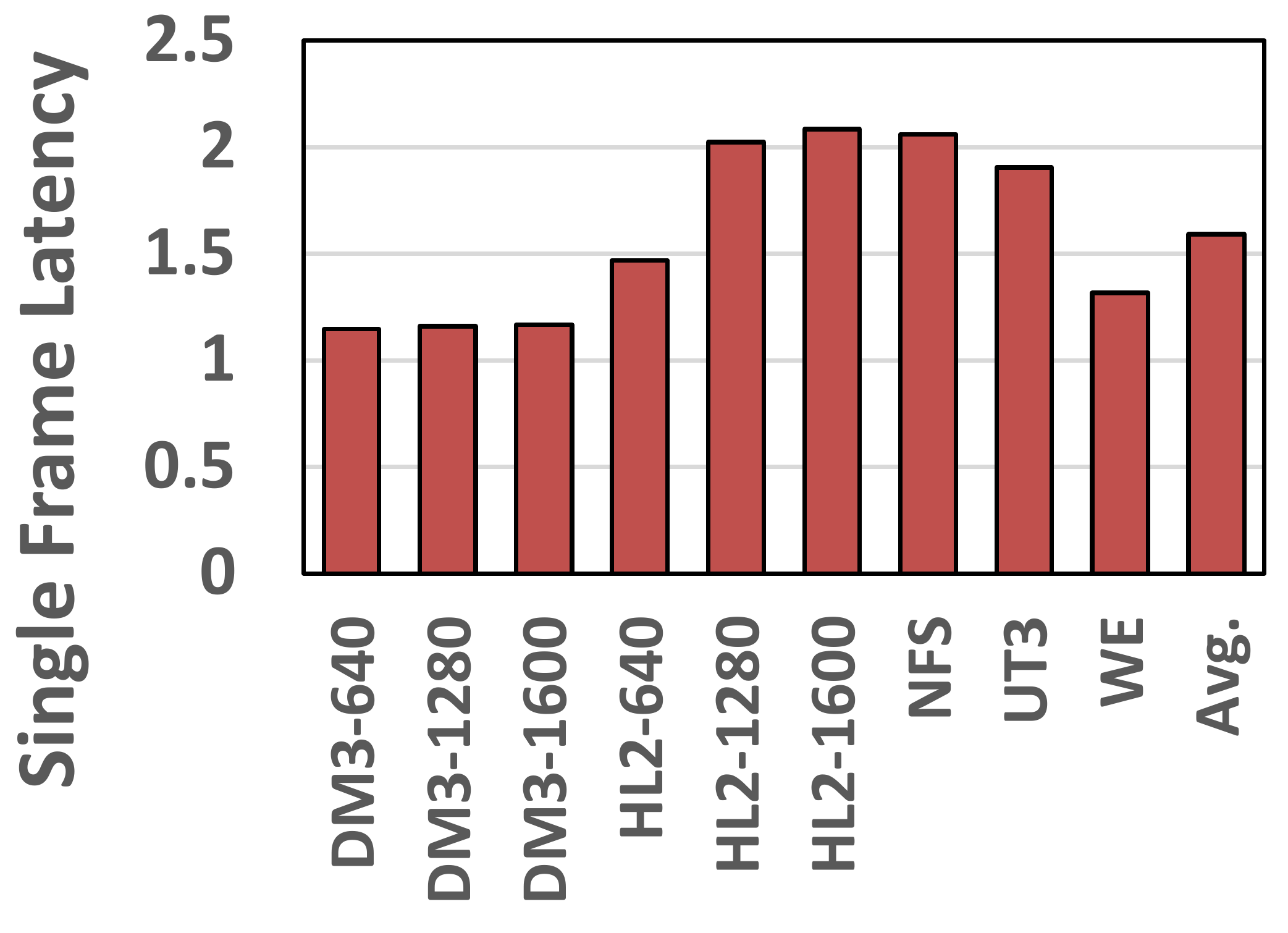}
 		\vspace{-0.55cm}
 	\end{minipage}
 	\vspace{-0.2cm}
 	\caption{ Normalized performance improvement (left) and single frame latency (right) across different benchmarks.}
 	\label{fig:AFR}
 	\vspace{-0.3cm}
 \end{figure}
 
\section{Characterizing Parallel Rendering Schemes on Future NUMA-based Multi-GPU Systems}
\label{sec:Character}

Aiming to reduce the NUMA-induced bottlenecks, a straightforward method is to employ parallel rendering schemes in VR applications to distribute a domain of graphics workloads on a targeted computing resource. While some parallel rendering frameworks such as Equalizer \cite{eilemann2009equalizer,eilemann2018equalizer} and OpenGL Multiple SDK \cite{OpenGLmultipipeSDK} have been used in many cluster-based PC games, the NUMA-based multi-GPU systems face some different challenges when performing parallel rendering. In this section, we perform a detailed analysis using three state-of-the-art parallel rendering schemes (including frame-level, tile-level and object-level parallel rendering) for VR application running on such future NUMA-based multi-GPU architectures, to further understand the design challenges.

\subsection{Alternate Frame Rendering (AFR)}
 Alternate Frame Rendering (AFR), also known as frame-level parallel rendering, executes one rendering process on each GPU in a multi-GPU environment. As Figure \ref{fig:AFRimplement} demonstrates, AFR distributes a sequence of rendering frames along with the required data across different GPMs. AFR is often considered to be a better fit for distributed memory environment since the separate memory spaces make the concurrent rendering of different frames easier to implement\cite{eilemann2018equalizer}. To separate our NUMA memory system into unique memory spaces, we leverage the software-level segmented memory allocation to reserve distributed memory segments for each frame. We also employ a simple task scheduler to map the rendering workloads of a frame to a specific GPM. The benefit of this design is to eliminate the inter-GPM commutation. 

Figure \ref{fig:AFR} shows the performance improvement and single frame latency affected by AFR scheme. The results are normalized to the baseline NUMA-based multi-GPU setup (with 64GB/s NVLink) where the entire system is viewed as a single GPU under the programming model and rendering workloads are directly launched to this system without specific parallel rendering scheduling. On average, AFR improves the performance (i.e., overall frame rate) by 1.67X comparing to the baseline setup. AFR not only eliminates the performance degradation of low bandwidth inter-GPU links, but also increases the rendering throughput by leveraging the SMP feature of the GPM. However, Figure \ref{fig:AFR}(right) also suggests that AFR increases the single frame latency by 59\% as a frame is processed by only one GPU. This increased single-frame latency may cause significant motion anomalies, including judder, lagging and sickness in VR system \cite{ZhangLuo,PIMVR} because it highly impacts whether the corresponding display on VR head-gear device can be in sync with the actual motion occurrence. Additionally, we observe that AFR near-linearly increases the memory bandwidth and capacity requirement according to the pre-allocate memory space for each frame. This decreases the maximum system memory capacity which directly limits the rendering resolution, texture details and perceived quality for different VR applications.  

\subsection{Tile-Level Split Frame Rendering (SFR)}
In contrast to AFR, split frame rendering (SFR) tends to reduce the single-frame latency by splitting a single frame into smaller rendering workloads and each GPM only responses to one group of workloads. Figure \ref{fig:TBR1} shows a tile-level SFR which splits the rendering frame into several pixel tiles in the screen space and distributes these sets of pixels across different GPMs. This basic method is widely used in cluster-based PC gaming because it requires very low software effort\cite{vlachos2016advanced}. To employ tile-level SFR, we simply leverage the sort-first algorithm to define the tile-window size before the rendering tasks is processed in GPMs. Although this design can effectively reduce single-frame latency, its vertical pixel stripping \cite{vlachos2016advanced} does distribute left and right views into different GPMs, ignoring the redundancy of the two views. Thus, to enable the SMP under this tile-level SFR, an alternative is to employ a horizontal culling method, shown in Figure \ref{fig:TBR2}. It groups the left and right views as a large pixel tile so that the rendering workloads in the left view can by re-projected into the right via the SMP engine to reduce redundancy geometry processing and improve data sharing.  
  
\begin{figure}[t]
	\begin{center}
		\includegraphics[width=0.45\textwidth]{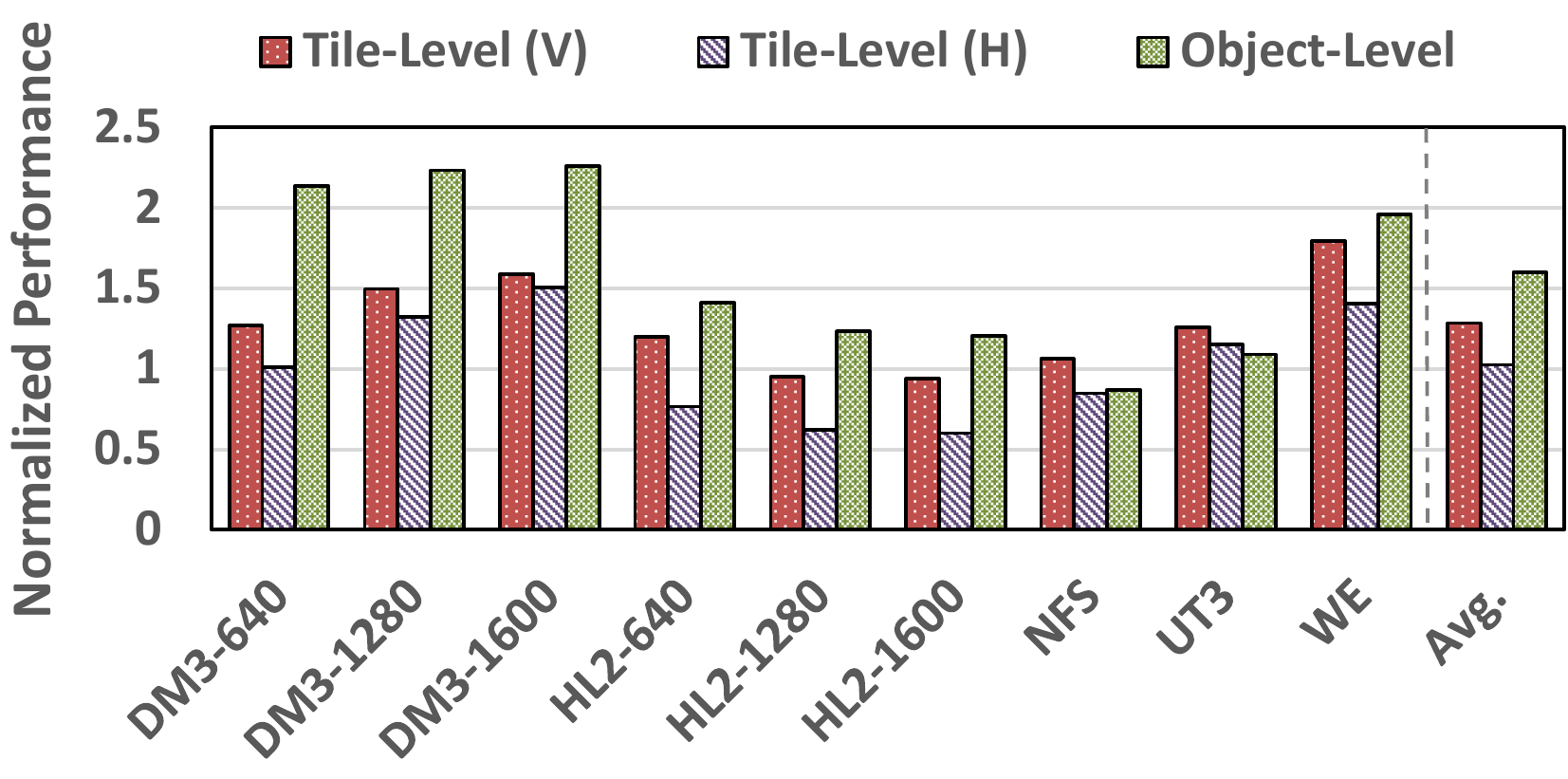}
		\vspace{-0.3cm}
		\caption{Normalized performance after enabling SFR across different benchmarks.}
		
		\label{fig:SFRperformance}
	\end{center}
	\vspace{-0.5cm}
\end{figure}   

\subsection{Object-Level Split Frame Rendering}

Distributing objects among processing units represents a specific type of split rendering frame (SFR). Figure \ref{fig:OBR} shows an example of object-level SFR which is often referred as sort-last rendering \cite{eilemann2009equalizer}. In contrast to the traditional vertical and horizontal tile-level SFR, the distribution under object-level SFR begins after the GPU starts the rendering process. During object-level SFR, a root node is selected (e.g., GPM0 in this example) to distribute the rendering objects to other working units (e.g., GPM1, GPM2 and GPM3). Once a worker completes the assembled object, the output color in its local DRAM is sent to the root node to composite the final frame. In this study, we first profile the entire rendering process to get the total number of rendering objects, and then issue them to different GPMs in a round-robin fashion. Note that only one object is executed in each GPM at a time for better data locality. Although this object distribution can also occur during rendering process (e.g., between rasterization and fragment processing \cite{Kim2017}), it typical requires to insert additional inter-GPM synchronization which may cause increasing inter-GPM traffic and performance degradation. Thus, we only distribute the objects at the beginning of the rendering pipeline for our experiments.

\begin{figure}[t]
	\begin{center}
		\includegraphics[width=0.45\textwidth]{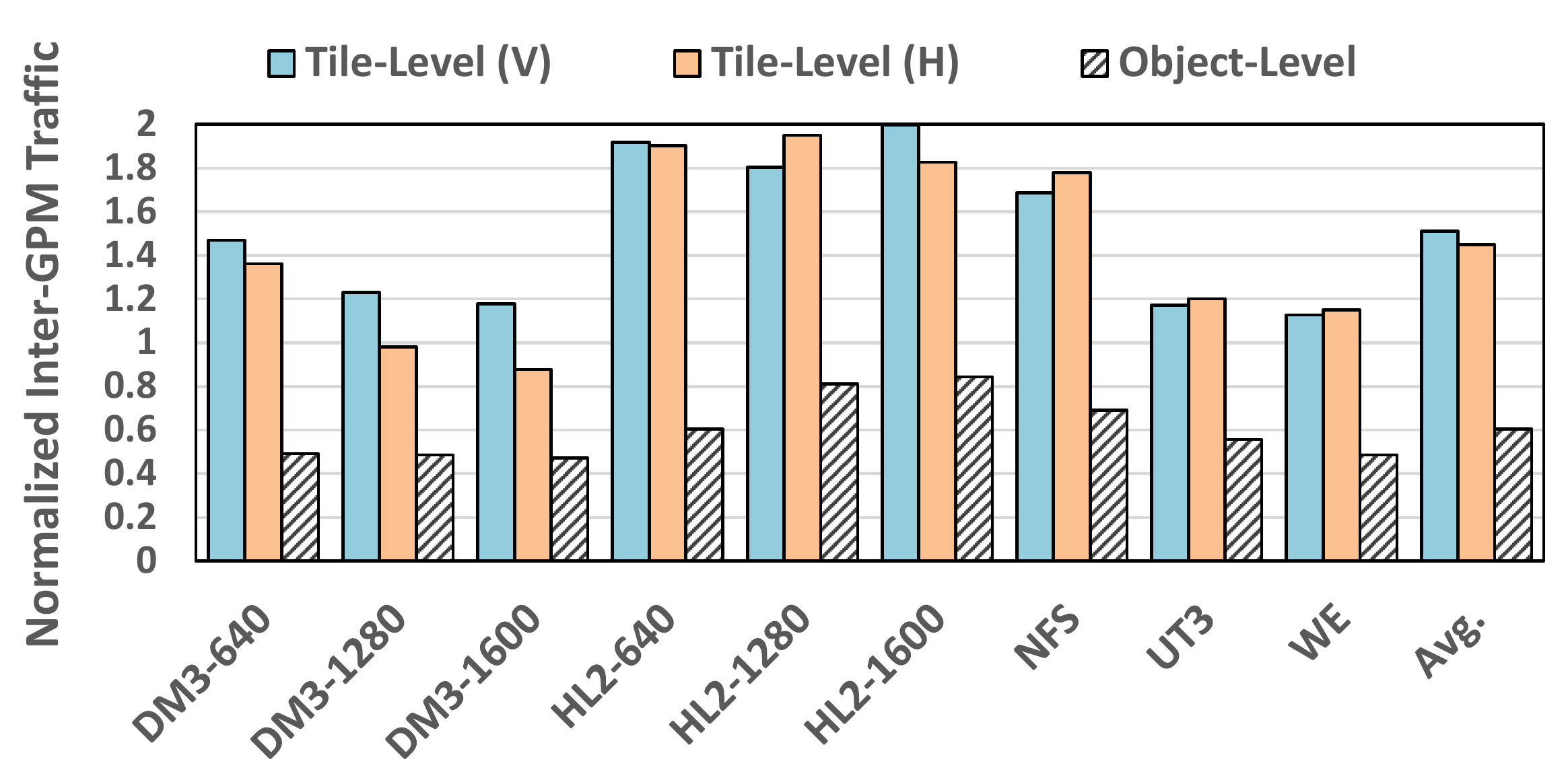}
		\vspace{-0.3cm}
		\caption{Total inter-GPM memory traffic across different benchmarks.}
		
		\label{fig:GPMBW}
	\end{center}
	\vspace{-0.3cm}
\end{figure}

Figure \ref{fig:SFRperformance} and \ref{fig:GPMBW} illustrate the performance (i.e., the overall frame rate) impact and inter-GPM memory traffic for different SFR scenarios. The results are normalized to the baseline setup. We have the following observations:

(i) The tile-level SFR schemes only slightly improve the rendering performance over the baseline case, e.g., on average 28\% and 3\% for Tile-level (V) and Tile-level (H), respectively. This is because although processing a small set of pixels via tile-level SFR can improve the data locality within one GPU, the tile-level SFR schemes increase the inter-GPM memory traffic by an average of 50\% for the vertical culling (V) and 44\% for horizontal culling (H) due to the object overlapping across the tiles. While the horizontal culling (H) fails to capture the data sharing for large objects (e.g., the bridge on the right side of Figure \ref{fig:TBR2}), vertical culling (V) ignores the redundancy between the left and right view. Since when applying SMP-based VR rendering the GPMs do not render the left and right views simultaneously, the large texture data have to be moved frequently across the GPMs.

(ii) The object-level SFR outperforms tile-level SFR schemes and achieves an average of 60\%, 32\% and 57\% performance improvement over the baseline, tile-level (V) and tile-level (H), respectively. The speedups are mainly from the inter-GPMs traffic reduction, indicated by Figure \ref{fig:GPMBW}. By placing the required data in the local DRAM for the rendered objects, Object-level SFR reduces approximately 40\% of inter-GPMs traffic compared to the baseline. However, the state-of-the-art object-level SFR can not fully address the NUMA-induced performance bottlenecks for VR execution, because it still executes the objects from the left and right views separately. In other word, it ignores the multi-view redundancy in VR applications which limits its rendering efficiency. 

\begin{figure}[t]
	\begin{center}
		\includegraphics[width=0.4\textwidth]{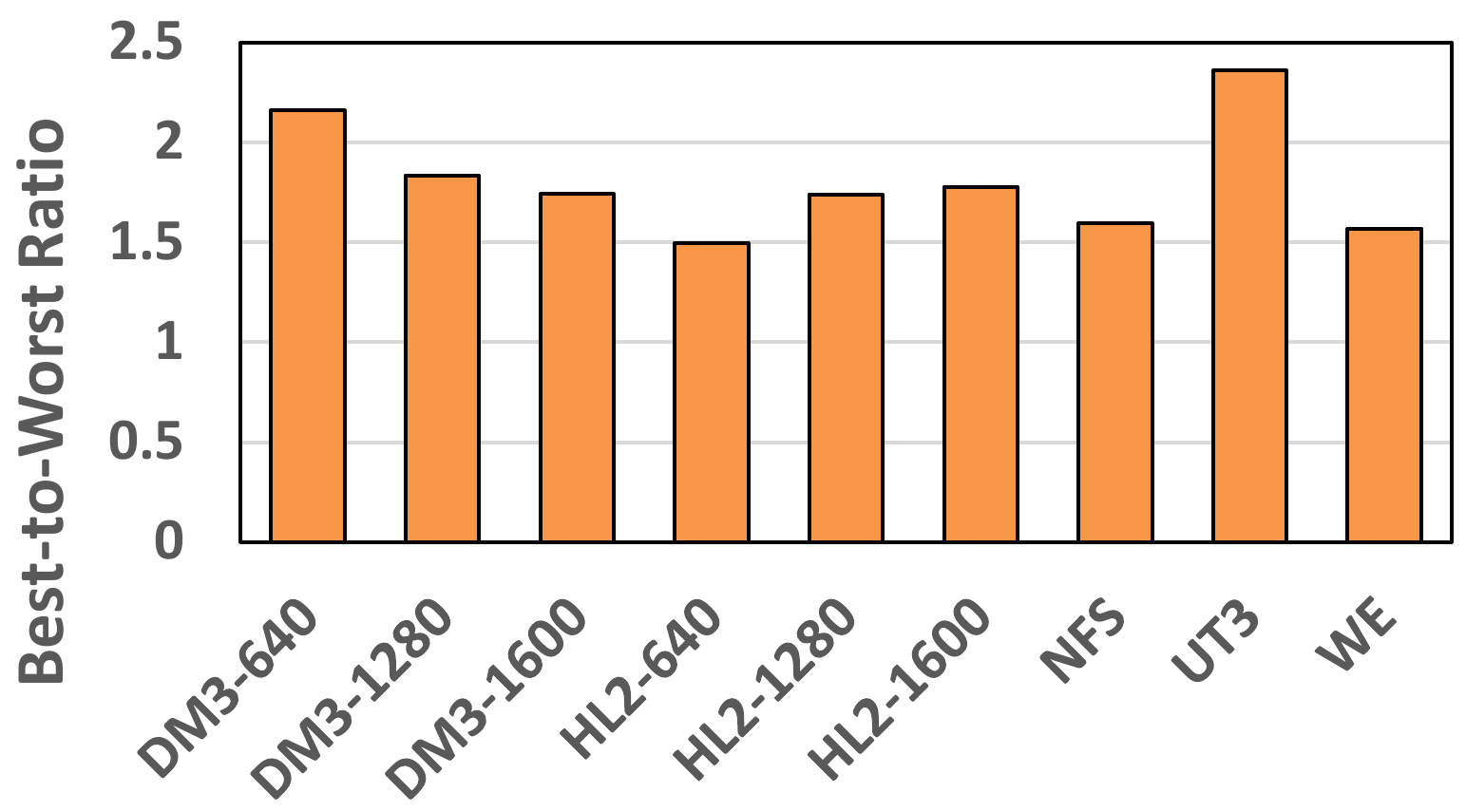}
		\vspace{-0.2cm}
		\caption{The best-to-worst performance ratio among GPMs in object-level SFR across different workloads.}
		
		\label{fig:BtoW}
	\end{center}
	\vspace{-0.3cm}
\end{figure} 

(iii) Additionally, we also observe that the object-level SFR is challenged by low load balance and high composition overhead. Figure \ref{fig:BtoW} shows the ratio between the best and the worst performance among different GPMs under the Round-Robin object scheduling policy. Since each object has a variety of graphical properties (e.g., the total amount of triangles, the level of details, the viewport window size, etc), the processing time is typically different for each object. If one GPM is assigned more complex objects than the others, it will take more time to complete the rendering tasks. Since the overall performance of Multi-GPU system is determined by the worst-case GPM processing, low load balance will significantly degrade the overall performance. Meanwhile, the high composition overhead (i.e., assembling all the rendering outputs from different GPMs into a frame) also contributes to the unbalanced execution time. As we mentioned previously, only the root node is employed to distribute and composite rendering tasks in the current object-level SFR. In this case, extra workloads will be issued to the root node while the ROP units of the other GPMs can not be fully utilized during this color output stage, causing bad composition scalability \cite{OpenGLmultipipeSDK}. Therefore, we aim to propose software-hardware support to efficiently handle these challenges facing the state-of-the-art object-level SFR in a NUMA-based multi-GPU environment. 
\section{Object-Oriented VR rendering f-ramework}
\begin{figure}[t]
	\begin{center}
		\includegraphics[width=0.45\textwidth]{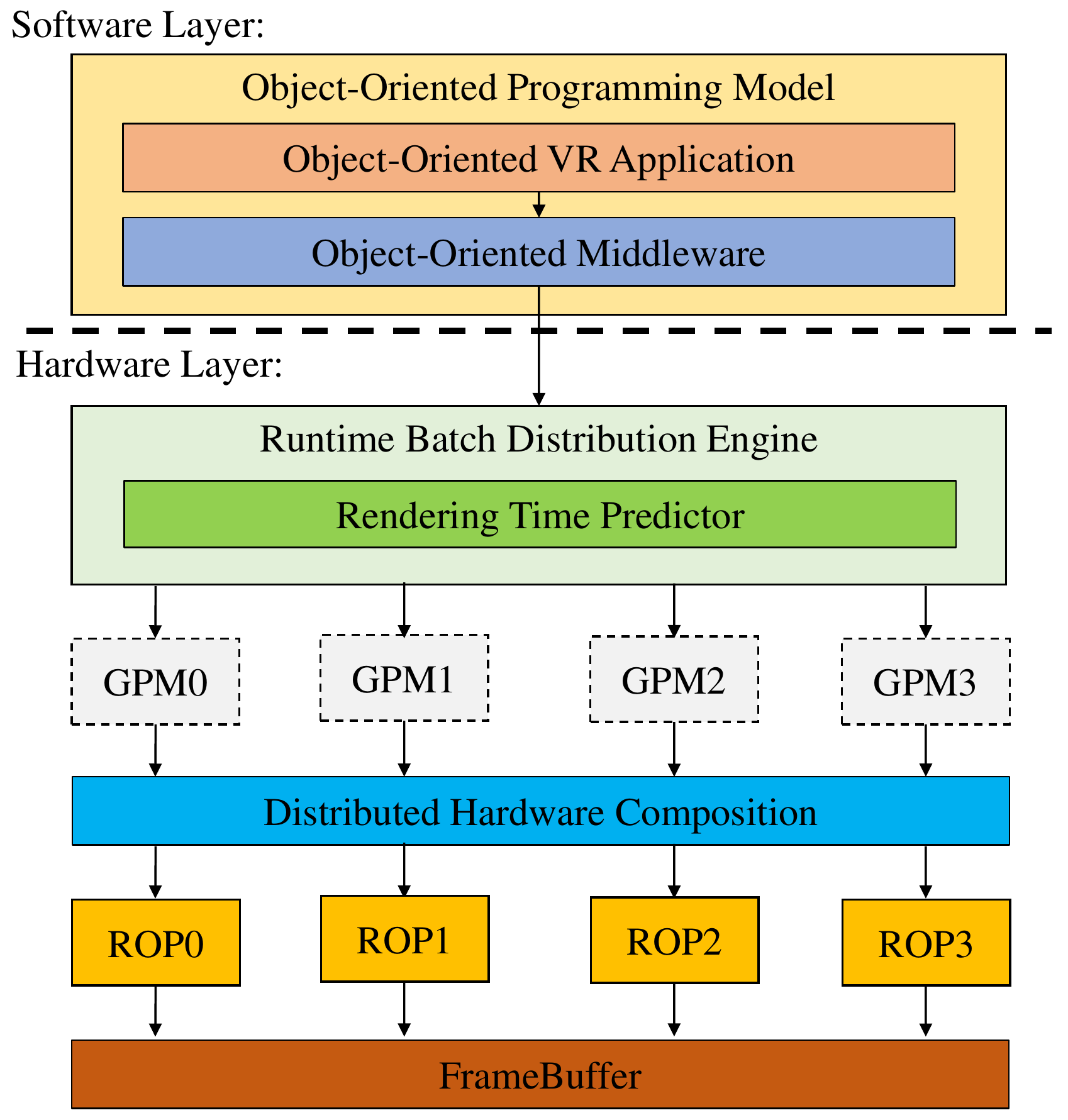}
		\vspace{-0.3cm}
		\caption{Our proposed object-oriented VR rendering framework (OO-VR).}
		
		\label{fig:OOVR}
	\end{center}
	\vspace{-0.5cm}
\end{figure}

In order to address the performance issues of the object-level SFR applied on future NUMA-based multi-GPU systems, we propose the object-oriented VR rendering framework (OO-VR). The basic design diagram is shown in Figure \ref{fig:OOVR}. It consists of several novel components. First, we propose an object-oriented VR programming model at the software layer to support multi-view rendering for the object-level SFR. It also provides an interface to connect the VR applications to the underlying multi-GPU platform. Second, we propose an object-aware runtime distribution engine at the hardware layer to balance the rendering workloads across GPMs. In OO-VR, this distribution engine predicts the rendering time for each object before it is distributed. It replaces the master-slave structure among GPMs so that the actual distribution is only determined by the rendering process. Finally, we design a distributed hardware composition unit to utilize all the ROPs of the GPMs to assemble and update the final frame output from the framebuffer. Due to the NUMA feature, the framebuffer is distributed across all the GPMs instead of only one DRAM partition, so that it can provide 4x output bandwidth of the baseline scenario. We detail each of these components as follows. 

% since the object distribution is happened in the new scheduler instead of the master note,

\subsection{Object-Oriented VR Programming Model}

\begin{figure}[t]
	\begin{center}
		\includegraphics[width=0.48\textwidth]{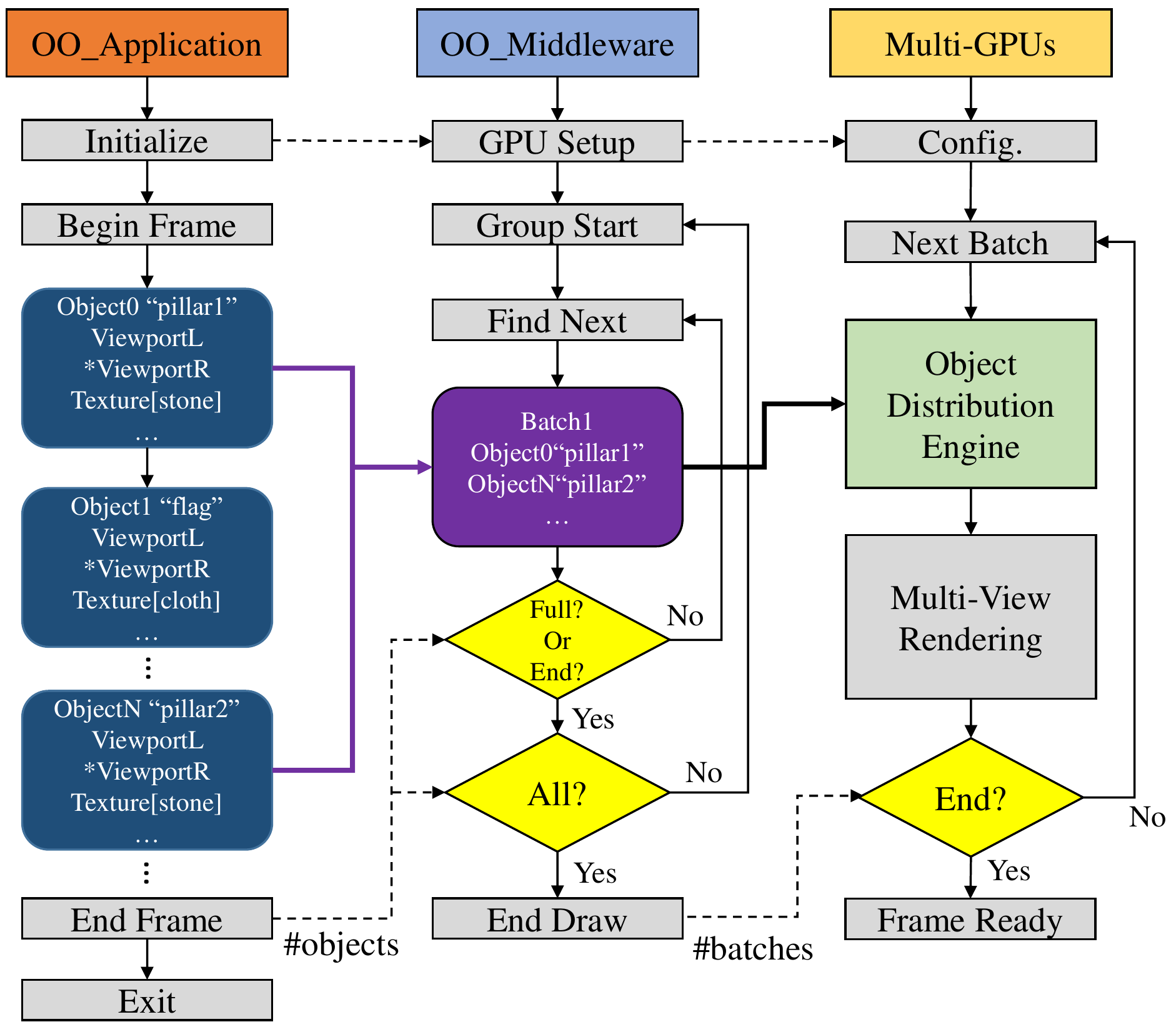}
		\vspace{-0.5cm}
		\caption{Simplified working flow of an Object-Oriented application, middleware and Multi-GPU rendering pipeline.}
		\label{fig:programmingmodel}
	\end{center}
	\vspace{-0.5cm}
\end{figure}

The Object-Oriented VR Programming Model extends the conventional object-level SFR as we introduced in Section \ref{sec:Character} and uses a similar software structure as today's Equalizer\cite{eilemann2009equalizer,eilemann2018equalizer} and OpenGL Multipipe SDK (MPK)\cite{OpenGLmultipipeSDK}. Figure \ref{fig:programmingmodel} uses an simplified working flow diagram to explain our programming model. In this study, we propose two major components that drive our OO-VR programming model: \textit{Object-Oriented Application (OO\_Application)} to drive the VR multi-view rendering for each object, and \textit{Object-Oriented Middleware (OO\_Middleware)} to reorder objects and regroup the ones that share similar features as a large batch which acts as the smallest scheduling units on the NUMA-based multi-GPU system. 

\textbf{The OO\_Application} provides a software interface (dark blue box) for developers to merge the left and right views of same object as a single rendering tasks. The OO\_Application is designed by extending the conventional object-level SFR. For each object, we replace the original viewport which is set during the rendering initialization with two new parameters -- \textit{viewportL} and \textit{viewportR}, each of which points to one view of the object. In order to enable rendering multi-views at the same time, we apply the built-in openGL extension $GL\_OVR\_multiview2$ to set two viewports ID for a single object. After that, each SMP engine integrated in a GPM automatically renders the left and right views to its own positioning using the same texture data. We also design an auto-model to extend the conventional object-level SFR to enable multi-view rendering through generating two fixed viewports for each object via shifting the original viewport along the X coordinate. In this case, only one rendering process needs to be setup for each object. In constrast to the single-path stereo rendering enabled in modern VR SDKs \cite{VRworks,eilemann2009equalizer}, our OO\_Application does not decompose the left and right views during the rendering initialization so that it still follows the execution model of the object-level SFR.  
 
\textbf{OO\_Middleware} is the software-hardware bridge to connect the OO\_Application and multi-GPU system. It is automatically executed during the application initialization stage to issue a group of the objects to the rendering queue of the multi-GPU system. In the conventional object-level SFR, the objects are scheduled in a master-slave fashion following the programmer-defined order. However, different objects that may share some common texture data are not rendered on the same GPM. As Figure \ref{fig:programmingmodel} illustrates, both "pillar1" and "pillar2" share the common "stone" texture. If they are rendered on different GPMs, the "stone" texture may need to be reallocated, increasing remote GPM access. In OO-VR, we leverage OO\_Middleware to group objects based on their \textit{texture sharing level} (TSL) to exploit the data locality across different objects. 

To implement this, OO\_Middeware first picks an object from the head of the queue as the root. It then finds the next independent object of the root and computes the TSL between the two using Equation (\ref{eq:TSL}).
\begin{equation}\footnotesize
    TSL =  \sum_{t}^{T} \left(P_{r}\left(t\right) \cdot P_{n}\left(t\right)\right) / \sum_{t}^{T}P_{r}\left(t\right)
\label{eq:TSL}
\end{equation}    
Where $t$ is the shared texture data between the two objects, $P_{r}(t)$ and $P_{n}(t)$ represent the percentages of $t$ among all the required textures for the root and the target object.

TSL represents how many texture data will be shared if we group the target object with the root. If TSL is greater than 0.5, we group them together as a batch and this batch then becomes the new root which consists all textures from the previous iterator and the target object. After this grouping, the OO\_Middleware removes the target object from the queue and continues to search for the next object until the total number of triangles within the batch is higher than 4096, or all the objects in the queue have been selected. The triangle number limitation is used to prevent load imbalance from an inflated batch. 
%However, we further design a batch decomposing method within the object distribute engine so that the triangle limitation only slightly affects the overall performance. 

After this step, this batch is marked as ready and issued to a GPM in the system for rendering. Finally, the OO\_Middleware repeats this grouping process for all the objects in the frame until there is no object in the queue. Note that for the objects that have dependency on any of the objects in a batch, we directly merge them to the batch and increase the triangle limitation so that they can follow the programmer-defined rendering order.      

\subsection{Object-Aware Runtime Distribution Engine}

\begin{figure}[t]
	\begin{center}
		\includegraphics[width=0.43\textwidth]{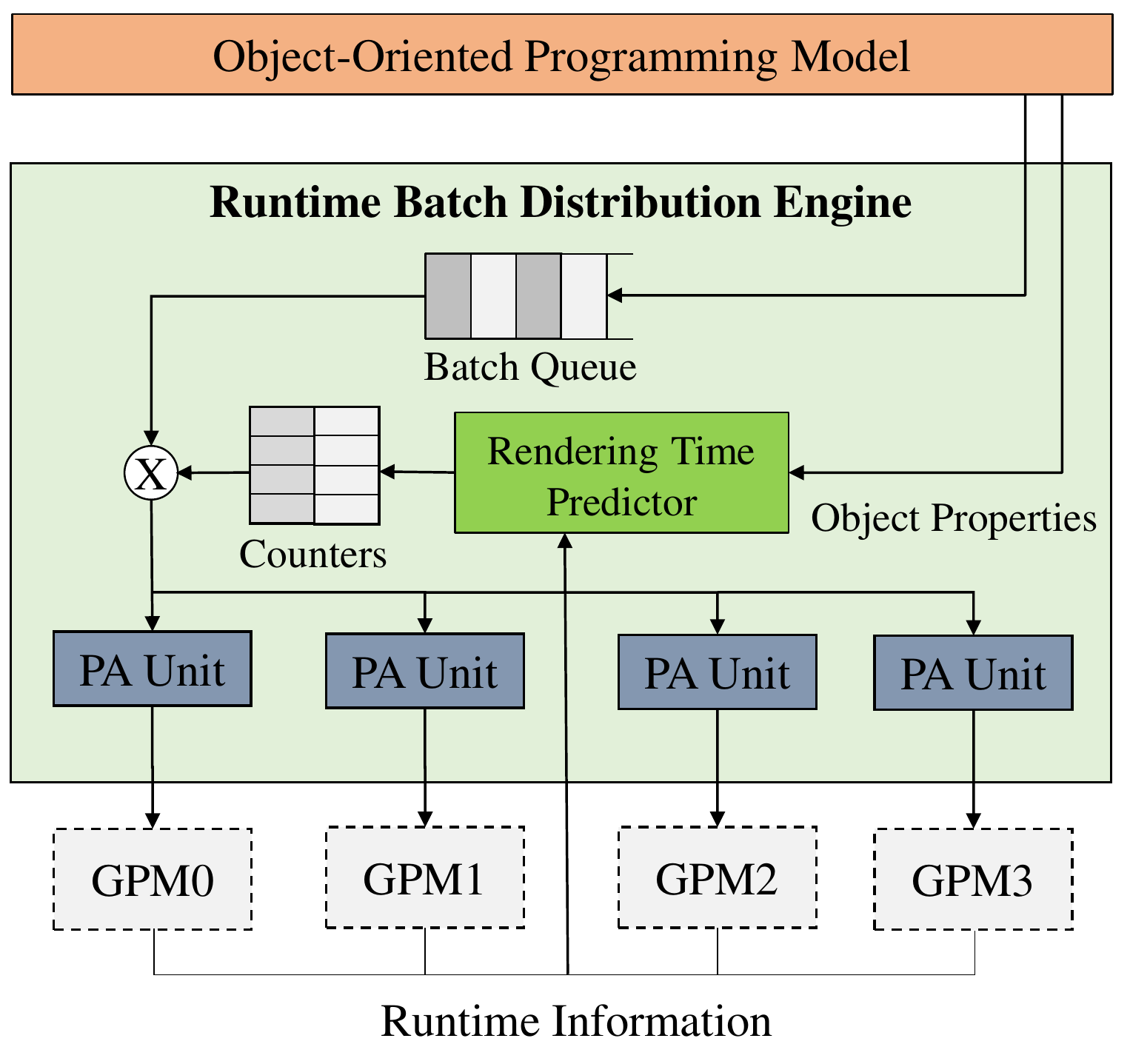}
		\vspace{-0.4cm}
		\caption{The architecture design of the object-aware runtime distribution engine.}
		
		\label{fig:distributeEngine}
	\end{center}
	\vspace{-0.7cm}
\end{figure}

After receiving the batches from the OO\_Middleware, the Multi-GPU system needs to distribute them across different GPMs for multi-view rendering. For workload balancing, we propose an object-aware runtime distribution engine at the hardware layer instead of using the software-based distribution method based on master-slave execution used in the conventional object-level SFR. Comparing to the software-level solution which needs to split the tasks before assigning to the multi-GPU system, the hardware engine provides efficient runtime workload distribution by collecting rendering information. Figure \ref{fig:distributeEngine} illustrates the architecture design of the proposed distribution engine. The new hardware architecture is implemented as a micro-controller for the multi-GPU system which responses for predicting the rendering time for each batch, allocating an object to the earliest available GPM, and pre-allocating memory data using the Pre-allocation Units (PAs) in each GPM. 

Recall the discussion in Section 3 for our evaluation baseline, we employ the first-touch memory mapping policy (FT) \cite{arunkumar2017mcm} to allocate the required data in the local DRAM. Although FT can help reduce inter-GPM memory access, it can also cause performance degradation if the required data is not ready during the rendering process. As a result, we consider to pre-allocate data before objects are being distributed across GPMs. In this case, OO-VR needs to be aware of the runtime information of each GPM to determine which GPM is likely to become idle the earliest. 
 
In order to obtain this information, we need to predict approximately how long the current batch will be completed. Equation (\ref{eq:predition}) shows a basic way to estimate the rendering time of the current task $X$, introduced by \cite{Wimmer}:  
\begin{equation}
t\left(X\right) = RT\left(g_x, c_x, HW, ST\right)
\label{eq:predition}
\end{equation}  
Where ${g_x,c_x}$ is the geometry and texture property of the object $X$, $HW$ is the hardware information, and $ST$ is the current rendering step (i.e., geometry process, multi-view projection, rasterazition or fragment process) of the object $X$.  

While a complex equation can increase the estimation accuracy, it also requires more comprehensive data and increases hardware design complexity and computing overhead. Because the objective of our prediction model is to identify the earliest available GPM instead of accurately predicting each batch's rendering time, 
we propose a simple linear memorization-based model to estimate the rendering time as Equation (\ref{eq:RTpredition}):
\begin{equation}
t\left ( X \right ) = c_0 \cdot \#triangle_x = c_1 \cdot \#tv_x + c_2 \cdot \#pixel_x
\label{eq:RTpredition}
\end{equation}   
Where $\#triangle_x$, $\#tv_x$ and $\#pixel_x$ represent the triangle counts, the number of transformed vertexes and the number of rendered pixels of the current batch, respectively. $c_0$, $c_1$ and $c_2$ represent the triangle, vertex and pixel rate of the GPM.

After building this estimation model, we split the prediction process into two phases: total rendering time estimation and elapsed rendering time prediction. We setup two counters to record the total rendering time and the elapsed rendering time for each GPM. First, we leverage $\#triangle_x$ (which can be directly acquired from the OO\_Application) to predict the total rendering time. During rendering, the distribution engine tracks $\#tv_x$ and $\#pixel_x$ from GPMs to calculate the elapsed rendering time. If the $\#tv_x$ or $\#pixel_x$ increases by 1, the elapsed counter increases by $c_1$ or $c_2$, respectively. At the end, by comparing the distance between the two counters from each GPM, we can predict which GPM will become available first.

At the beginning of the rendering, the distribution engine uses the first 8 batches to initialize $c_0$, $c_1$ and $c_2$. The first 8 batches will be distributed across GPMs under the Round-Robin object scheduling policy and baseline FT memory mapping scheme is also applied to allocate the rendering data. After GPMs complete this round of 8 batches, the total rendering time will be sent back to the distribution engine to calculate $c_0$, $c_1$ and $c_2$. Then, starting from the 9th batch, the rendering time predictor is enabled to find the earliest available GPM. After that, the PA Unit pre-allocates the required data to the selected GPMs, and the rendering time predictor updates the predicted total rendering time by increasing the triangle counts. Note that we limit the maximum size of the batch queue to 4 objects to reduce the memory space requirement. Multiple batches could be distributed onto one GPM at the same time. In this case, a PA Unit sequentially fetches the data based on the order of the batch ID. 

We further observe that even though distribution engine can effectively balance the rendering tasks, it is possible that some large objects may still become the performance bottleneck if all the other batches have been completed. To fully utilize the computing and memory resources of these idle GPMs, we employ a simple fine-grained task mapping mechanism to fairly distribute the rest of the processing units (e.g. triangles in geometry process and fragments in fragment process) to idle GPMs based on their IDs. Meanwhile, the PA units duplicate the required data to the corresponding unused DRAM to eliminate inter-GPMs access for these left-over fine-grained tasks.    

\subsection{Distributed Hardware Composition Unit}

\begin{figure}[t]
	\begin{center}
		\includegraphics[width=0.45\textwidth]{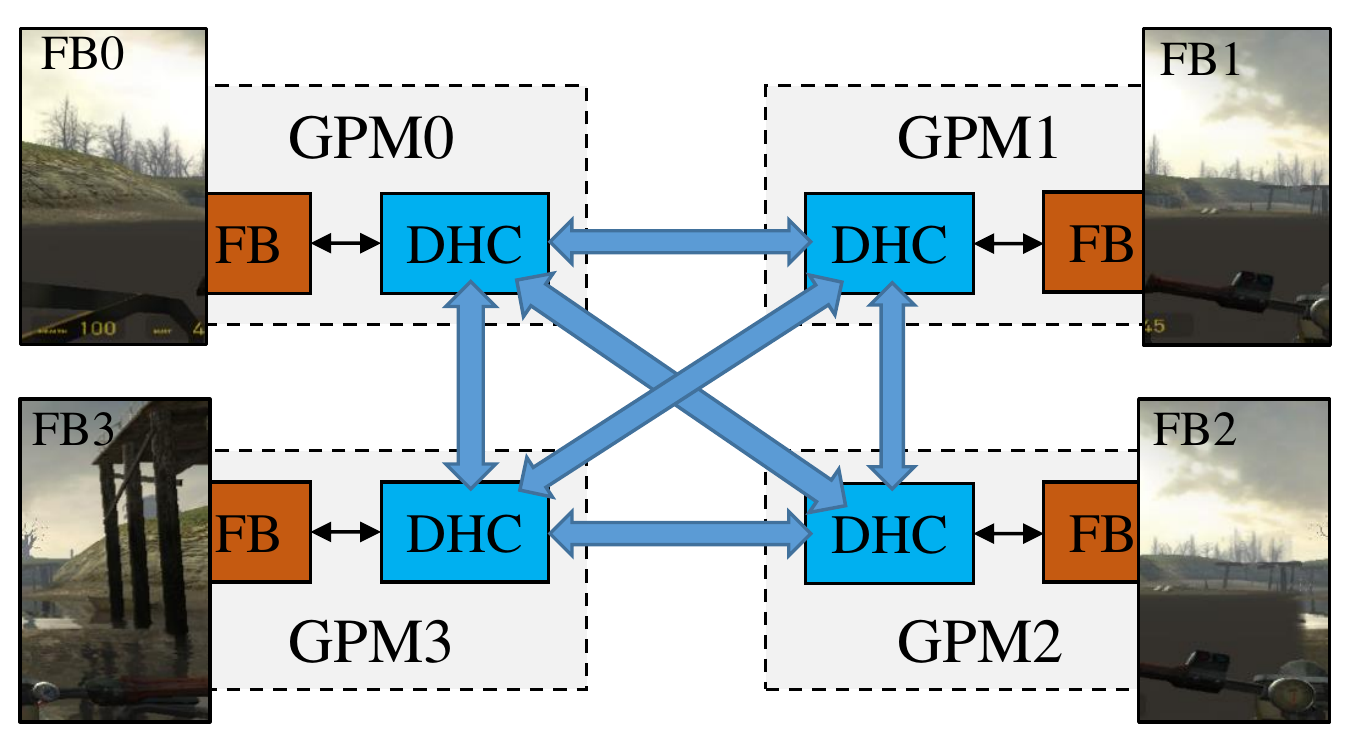}
		\vspace{-0.4cm}
		\caption{The distributed FrameBuffer for hardware composition.}
		
		\label{fig:DHC}
	\end{center}
	\vspace{-0.5cm}
\end{figure}

In the conventional object-level SFR, the entire FrameBuffer (FB) is mapped in the DRAM of the master node, and all the rendering outputs will then be transmitted to the master node for the final composition. Although the color outputs can be executed asynchronously with the shader process, a small amount of ROPs in a single GPM limits the pixel rate which impacts the overall rendering performance. Since the NUMA-based multi-GPU system can be considered as a large single GPU, we consider to distribute the composition tasks across all the GPMs which is currently not supported due to the lack of relevant communication mechanism and hardware.     

For example, shown in Figure \ref{fig:DHC}, we first split the entire FB into 4 partitions using the screen-space coordinate of the final frame. Here we employ the same memory mapping policy as the vertical Tile-level SFR (V). Based on this, we propose the distributed hardware composition unit (DHC) to determine which part of FrameBuffer is used to store what color outputs of the final frame. This design is based on the observation that the color outputs of the final frames only incur a small number of memory access compared to the main rendering phase so that the small amount of remote communication for this phase will not become a performance bottleneck for NUMA-based multi-GPU systems. This is also why vertical culling shown in Figure \ref{fig:DHC} can perform well as the last stage of VR rendering (i.e., after the object-aware runtime distribution for the main rendering phase) since the inter-GPM bandwidth can be effectively utilized by the distributed hardware composition.

%It checks the pixel coordinate of output color and generates memory address for it. If the pixel coordinate locates in the local FrameBuffer, the output color will be sent to the ROPs within Local GPM. In otherwise, the output color is distributed to remote GPMs via inter-GPMs links. 
%Because the NUMA-based multi-GPU system exhibits continued memory space, DHC can directly leverage the inter-GPM Links to support the color communication. In addition, by leveraging the first two proposed technologies to eliminate the remote texture access, the entire inter-GPM bandwidth can be fully utilized by the distributed hardware composition. Meanwhile, previous works \cite{PIMrendering,Perception3D} already proved that  the color output only incur a small number of memory accesses so that it will not become the new bottleneck for NUMA-based multi-GPUs system. %

\subsection{Overhead Analysis}
The major hardware components added into the existing multi-GPU system is the object-aware runtime distribution engine, which consists of a rendering time predictor, GPM counters and a batch queue. For the baseline Multi-GPU architecture that we modeled for this work (Table 2), we allocate 64 bits for each counter and 16 bits for each batch ID to store the predicted rendering time. Additionally, to predict the total and elapsed rendering time, twelve 32-bits registers are used to track the triangle counts, the number of transformed vertexes and the number of the rendered pixels for the current batches. In total, we only require 960 bits for storage and several small logic units. %, which incur negligible hardware overhead compared to the entire Multi-GPU system modeled in Table 2. 
We use McPAT\cite{McPAT} to evaluate the area and energy overhead of the added storage and logic units for the distribution engine. The area overhead is 0.59 $mm^2$ under 24nm technology which is 0.18\% to modern GPUs (e.g., GTX1080). The power overhead is 0.3W which is 0.16\% of TDP to GTX1080.

\section{Evaluation}

We model the object-oriented VR rendering framework (OO-VR) by extending AITTILA-sim \cite{attila}. To get the object graphical properties (e.g., viewports, number of triangles and texture data), we profile the rendering-traces from our real-game benchmarks as shown in Table \ref{table:benchmarks}. Then in ATTILA-sim, we implement the OOVR programming model in its GPUDriver, and the object distribution engine in its command processor, and the distributed hardware composition during the color writing procedure. %The programming model automatically translates the original rendering tasks to OO\_application and groups the objects into large batches. \textcolor{red}{[you already explain how to translate task and group objects in the previous section, right?]} The distribution engine tracks the batch ID and the batch's graphical properties at runtime to schedule the batches in different GPMs. 
To evaluate the effectiveness of our proposed OO-VR design, we compare it with several design scenarios: (i) \textit{Baseline} - the baseline multi-GPU system with single programming model (Section \ref{sec:Background}); (ii) \textit{1TB/s-BW} - the baseline system with 1 TB/s inter-GPU link bandwidth; 
(iii) \textit{Object-level} - the Object-level SFR which distributes objects among GPMs (Section \ref{sec:Character}); (iv) \textit{Frame-level} - the AFR which renders entire frame within each GPM; and (v) \textit{OO\_APP} - the proposed object-oriented programming model (Section \ref{sec:Methodology}). We provide results and detailed analysis of our proposed design on performance, inter-GPU memory traffic, sensitivity study for inter-GPM link bandwidth and the performance scalability over the number of GPMs. 

\subsection{Effectiveness On Performance}  
    
\begin{figure}[t]
	\begin{center}
		\includegraphics[width=0.45\textwidth]{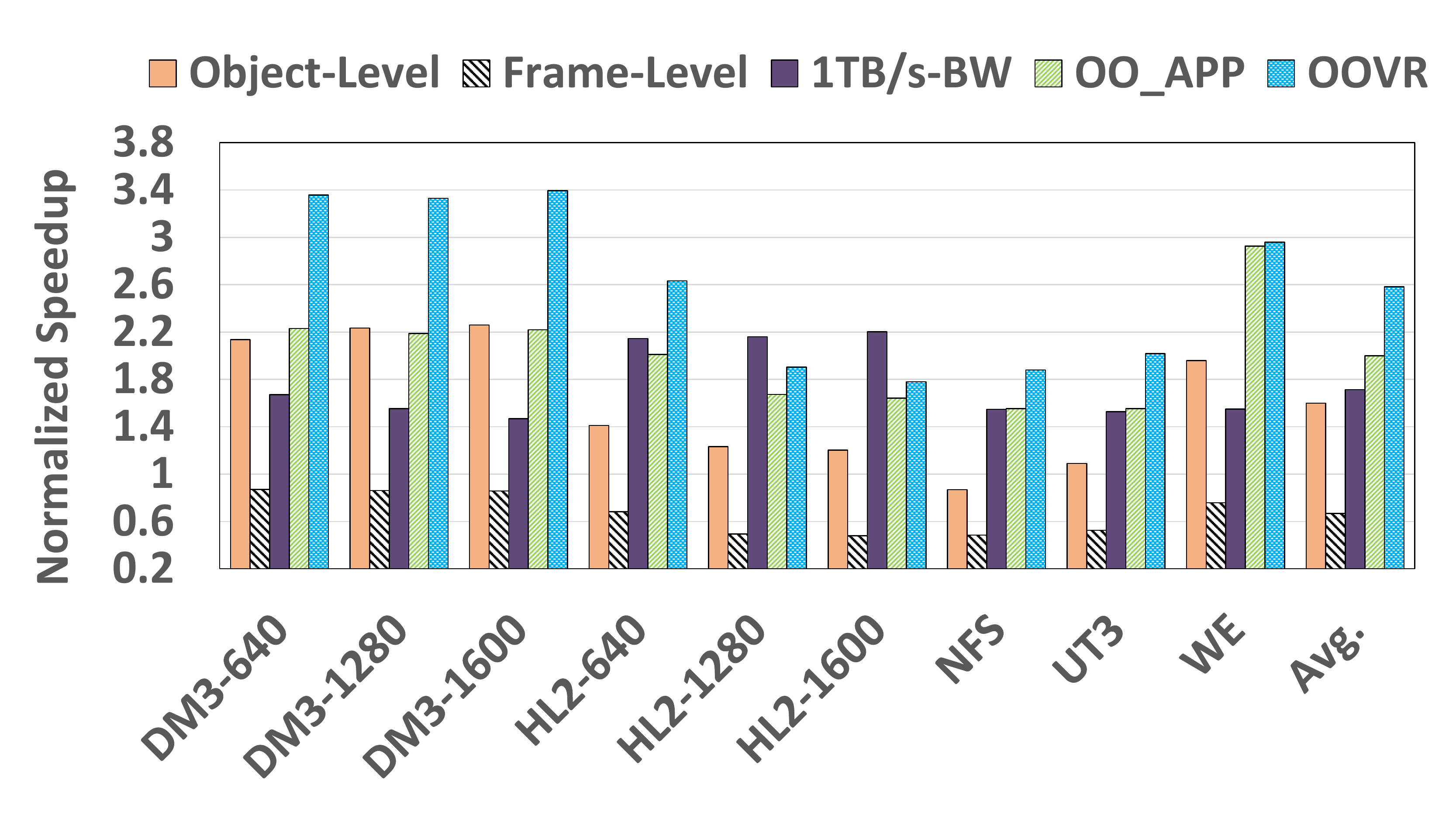}
		\vspace{-0.5cm}
		\caption{ Normalized speedup of overall VR rendering for single frame under different design scenarios.}
			\vspace{-0.5cm}
		\label{fig:result-performance}
	\end{center}
\end{figure} 

Fig.\ref{fig:result-performance} shows the performance results with respect to single frame latency under the five design scenarios. We gather the entire rendering cycles from the beginning to the end for each frame and normalized the performance speedup to baseline case. We show the performance speedup for single frame because it is critical to avoid motion sickness for VR. From the figure, we have several observations.

First, without hardware modifications, the OO\_APP improves the performance about 99\%, 39\% an 28\% on average comparing to the Baseline, Object-level SFR and 1TB/s-BW, respectively. It combines the two views of the same object and enable the multi-view rendering to share the texture data. In addition, by grouping objects into large batches, it further increases the data locality within one GPM to reduce the inter-GPM memory traffic. However, it still suffers serious workload unbalance. For instance, object-level SFR slightly outperforms OO\_APP when executing DM3-1280 and DM3-1600. This is because some batches within these two benchmarks require much longer rendering time than other batches, the software scheduling policy alone in OO\_APP can not balance the execution time across GPMs without runtime information.    
Second, we observe that on average, OO-VR outperforms Baseline, Object-level SFR and OO\_APP by 1.58x, 99\% and 59\%, respectively. With the software and hardware co-design, OO-VR distributes batches based on the predicted rendering time and provides better workload balance than OO\_APP. It also increases the pixel rate by fully utilizing the ROPs of all GPMs.

We also observe that OO-VR could achieve similar performance as Frame-level parallelism which is considered to provide ideal performance on overall rendering cycles for all frames (as shown in Fig.\ref{fig:AFR}(left)). However, in terms of the single frame latency, Frame-level parallelism suffers 40\% slowdown while OO-VR could significantly improve the performance.   
	
%for majority benchmarks except some benchmarks with very unbalanced batches (e.g., DM3, UT3). 
	
%Although OOVR fairly distributes the rendering tasks, the largest object in these benchmarks still becomes the bottleneck for rendering work. We believe object decomposing and fine-grain distribution methods could potentially solve this issue and leave it as our future research tasks.}     

%Note that we did not present the performance results with respect to the single frame latency because all the investigated techniques (except the Frame-Level parallelism technique) render single frame using the entire system, the overall rendering cycles is the product of the number of frames and single frame latency. In other words, the overall rendering cycles can well represent the single frame latency for these techniques. On the other hand, Frame-Level parallelism just uses one GPM to render one frame, it suffers 1.6X slowdown on single frame latency comparing to the baseline case (as shown in Fig.\ref{fig:AFR}(right)) although it provides the highest overall performance.    
 %shows the overall performance instead of single frame latency. For Frame-Level, that is assumed to provide the ideal performance in Multi-GPUs system, the single frame latency is about 1.6x slowdown comparing to the baseline case (as shown in Fig.\ref{fig:AFR}(right)). For other design scenarios, since they render single frame using entire system, the overall performance is equal to single frame latency. 

\subsection{Effectiveness On Inter-GPU Memory Traffic}

Reducing inter-GPM memory traffic is another important criteria to justify the effectiveness of OO-VR. Fig.\ref{fig:result-mem} shows the impact of OO-VR on inter-GPM memory traffic. Both Baseline and 1TB/s-BW have the same inter-GPM memory traffic, and Frame-Level is processing each frame in one GPM and has near-zero inter-GPM traffic. 
Moreover, the memory traffic reduction is mainly cause by our software-level design, the inter-GPM traffic is the same under the impact of OO\_APP and OO-VR. Therefore, Fig.\ref{fig:result-mem} only shows the results for Baseline, Object-Level and OO-VR, and we mainly investigate these three techniques in the following subsections. From the figure, we observe OO-VR can save 76\% and 36\% inter-GPM memory accesses comparing to the Baseline and Object-level SFR, respectively. This is because OO-VR allocates the required rendering data to the local DRAM of GPMs. The  majority inter-GPM memory accesses are contributed by the distributed hardware composition, command transmit and Z-test during fragment process. We observe that the delay caused by these inter-GPM memory accesses can be fully hidden by executing thousands of threads simultaneously in numerous shader cores.
In addition, the data transfer via the inter-GPM links also leads to higher power dissipation (e.g. 10pj/bit for board or 250pj/bit for nodes based on different integration technologies\cite{arunkumar2017mcm}). By reducing inter-GPM memory traffic, OO-VR also achieves significant energy and cost saving.    

\subsection{Sensitivity To Inter-GPM Link Bandwidth}

 \begin{figure*}[t]
	\centering
	\begin{minipage}[h]{0.37\textwidth}
		\includegraphics[width=\textwidth]{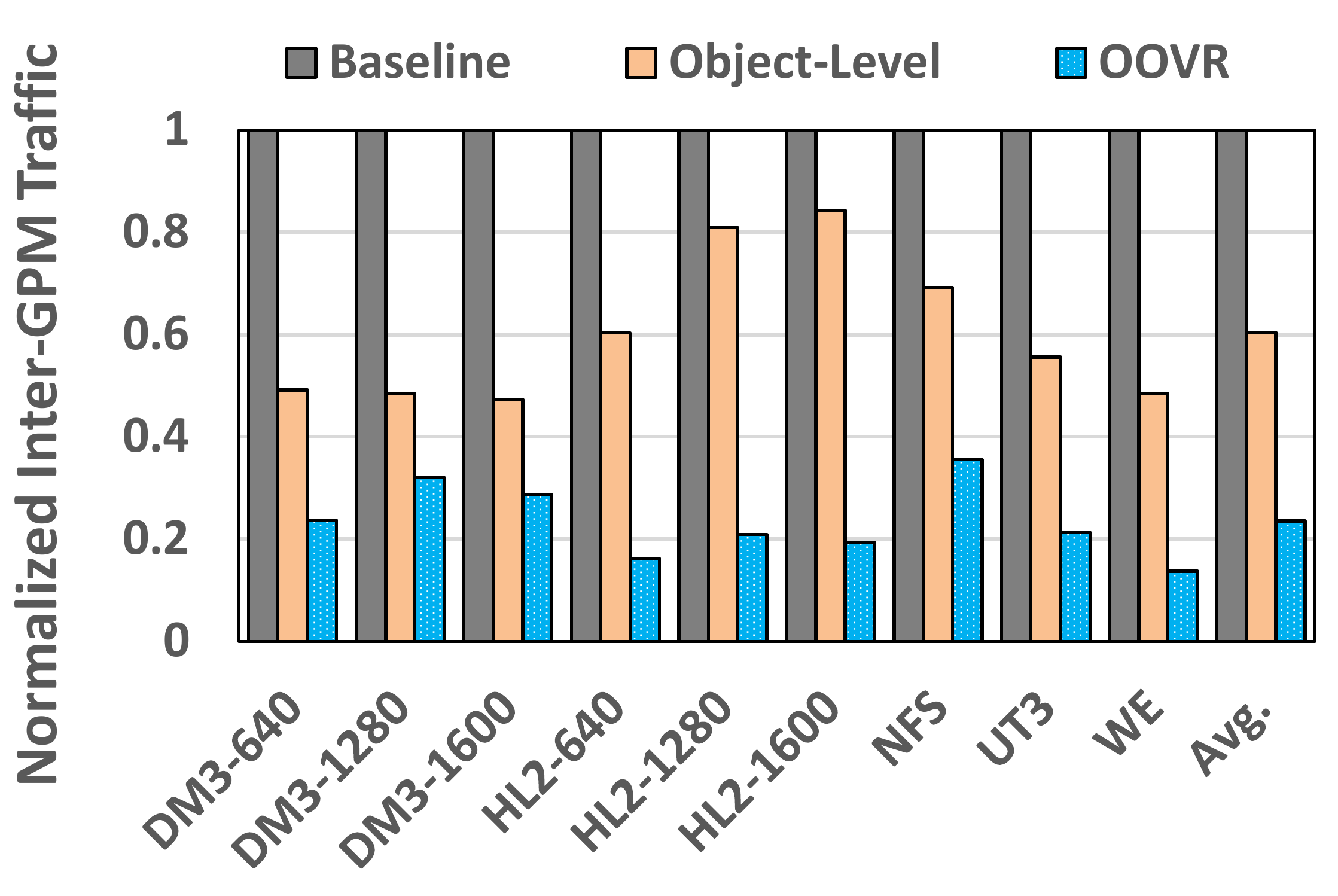}
	\vspace{-0.55cm}
	 \captionsetup{width=.95\linewidth}
	\caption{ Normalized inter-GPM memory traffic under different design scenarios.}
	
	\label{fig:result-mem}
	\end{minipage}
	\begin{minipage}[h]{0.3\textwidth}
		\includegraphics[width=1\textwidth]{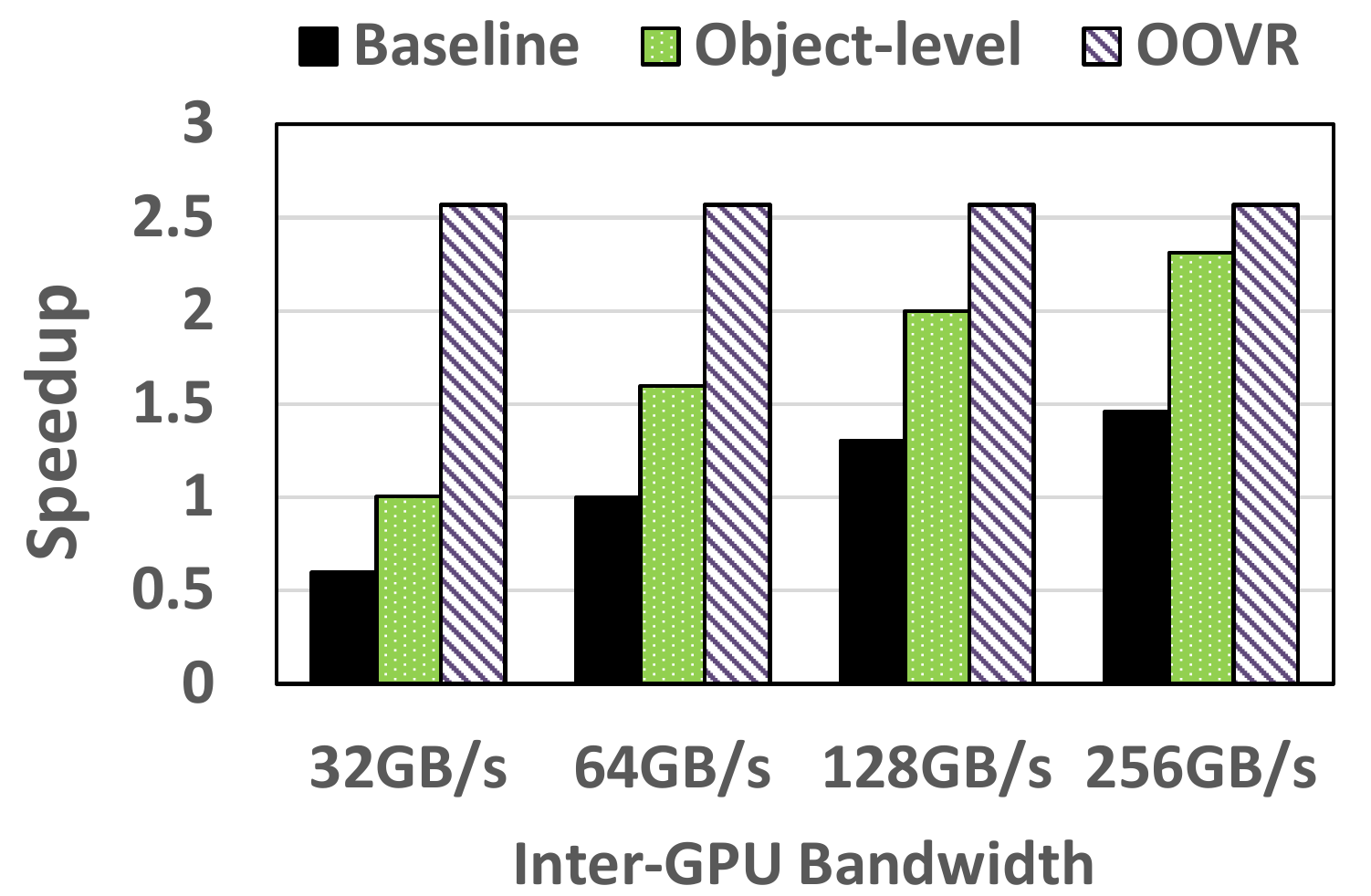}
		\vspace{-0.55cm}		
		 \captionsetup{width=.95\linewidth}
			\caption{Normalized speedup of the proposed OO-VR, object-level SFR and baseline under different inter-GPM link bandwidth.}
		\label{fig:memsensitive}
	\end{minipage}
	\begin{minipage}[h]{0.3\textwidth}
		\includegraphics[width=1\textwidth]{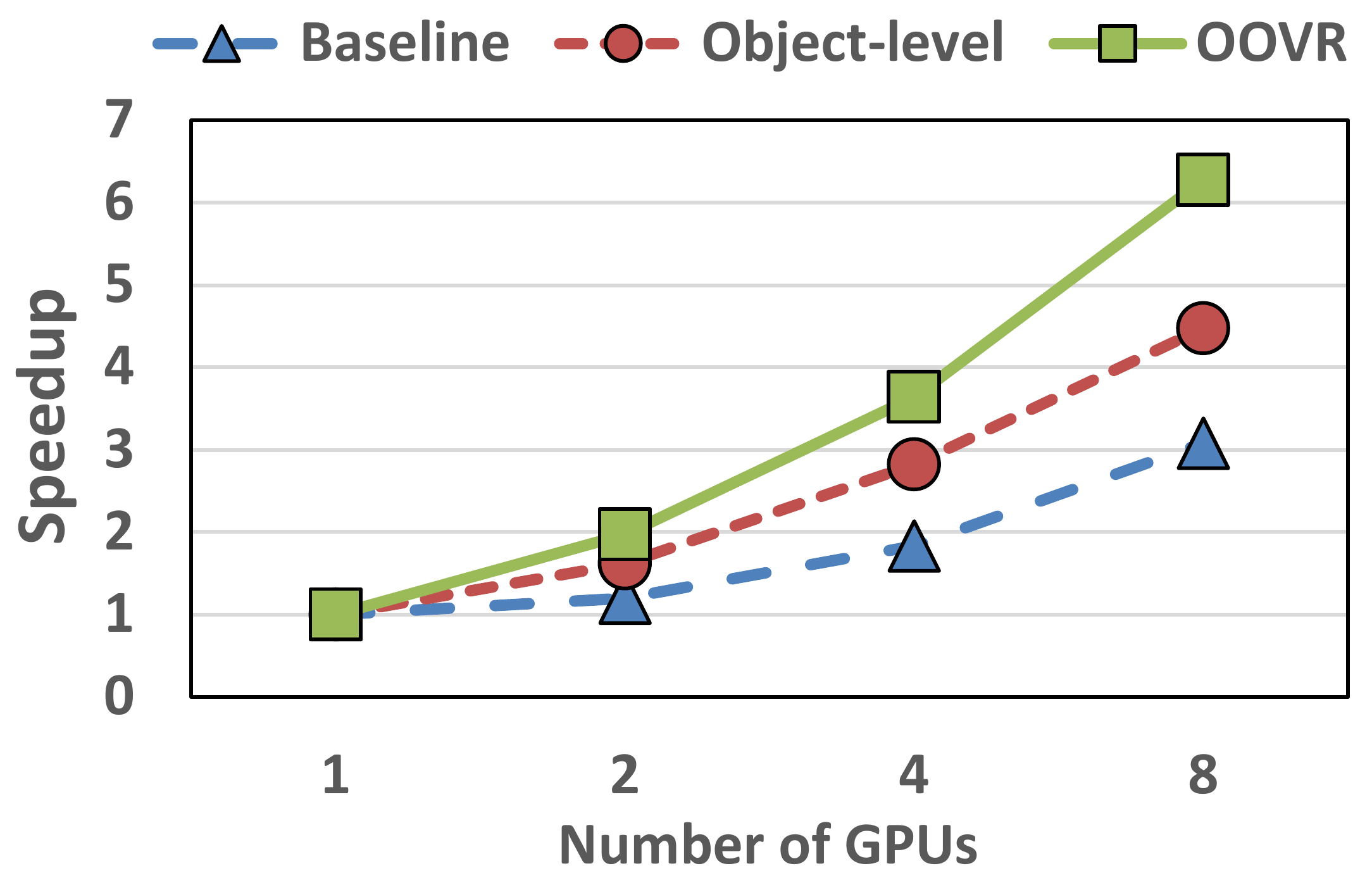}
		\vspace{-0.6cm}
		 \captionsetup{width=.95\linewidth}
		\caption{Normalized speedup of the proposed OO-VR, object-level SFR and baseline over single GPU under different number of GPMs.}
		\label{fig:numberscalability}
	\end{minipage}
	\vspace{-0.2cm}
\end{figure*}

Inter-GPU link bandwidth is one of the most important factors in multi-GPU systems. Previous works \cite{arunkumar2017mcm,milic2017beyond} have shown that increasing the bandwidth of inter-processor link is difficult and requires high fabric cost. To understand how inter-GPM link bandwidth impacts the design choice, we examine the performance gain of OO-VR under a variety of choices on link bandwidth. Fig.\ref{fig:memsensitive} shows the speedup under different link bandwidth when applying Baseline, Object-level SFR and our proposed OO-VR. In this figure, we normalize the performance to the Baseline with 64GB/s inter-GPM link. 
We observe that the inter-GPU link bandwidth highly affects the Baseline and Object-level SFR design scenarios. This is because these two designs cannot capture the data locality within the GPM to minimize the inter-GPU memory accesses during rendering. The large amount of shared data across GPMs significantly stalls the rendering performance. In the contrast, OO-VR fairly distributes the rendering workloads into different GPMs and convert numerous remote data to local data. By doing this, it fully utilizes the high-speed local memory bandwidth and is insensitive to the bandwidth of inter-GPM link even the inter-GPM memory accesses are not entirely eliminated. As the local memory bandwidth scales in future GPU design (e.g. High-Bandwidth Memory (HBM)\cite{HBM}), the performance of the future multi-GPU scenario is more likely to be constrained by inter-GPU memory. In this case, 
%ratio between local memory bandwidth to inter-GPM memory bandwidth may also increase. 
we consider the OO-VR can potentially benefit the future multi-GPU scenario by reducing inter-GPM memory traffic.
%with ever increasing asymmetric bandwidth between local and inter-GPU memory.  

\subsection{Scalability of OO-VR}
Fig.\ref{fig:numberscalability} shows the average speedup of the Baseline, Object-level SFR and OO-VR as the number of GPMs increases. The results are normalized to single-GPU system. %We configure the multi-GPU system by changing the number of GPMs to evaluate the scalability.
As the figure shows, the Baseline and Object-level SFR suffer limited performance scalability due to the NUMA bottleneck. With 8 GPMs, the Baseline and Object-level SFR only improve the overall performance by 2.08x and 3.47x on average over the single GPU processing. On the other hand, the OO-VR provides scalable performance improvement by distributing independent rendering tasks to each GPM. Hence, with 4 and 8 GPMs, it achieves 3.64x and 6.27x speedup over the single GPU processing, respectively. %However, we observe that increasing the number of GPMs also exacerbates unbalance problem among objects. In this case, cooperatively rendering among GPMs and data duplication mechanism may be employed to achieve continuous scalability for larger multi-GPUs system which tends to have high GPU memory capacity.    

\section{Related Work}

\textbf{Architecture Approach For NUMA Based Multi-GPU System.}
There have been many works \cite{Kim2017,arunkumar2017mcm,milic2017beyond} improving the performance for NUMA based multi-GPU system. Some of them\cite{arunkumar2017mcm, milic2017beyond,Vinson2018remote} introduce architectural optimizations to reduce the inter-GPM memory traffic for GPGPU application while Kim et al.\cite{Kim2017} redistribute primitives to each GPM to improve the scalability on performance. However, none of them discusses the data sharing feature of VR application. Our approach exploits the data locality during VR rendering to reduce the inter-GPM memory traffic and achieves scalable performance for multi-GPU system. 

\textbf{Parallel Rendering.} Currently, PC clusters are broadly used to render high-interactive graphics applications. To drive the cluster, software-level parallel rendering frameworks such as OpenGL Multipipe SDK \cite{OpenGLmultipipeSDK}, Chromium \cite{humphreys2002chromium}, Equalizer\cite{eilemann2009equalizer,eilemann2018equalizer} have been developed. 
They provides the application programming interface (API) to develop parallel graphics applications for a wide rand of platforms. Their works tend to split the rendering tasks during application development under different configurations.
In our work, we propose a software and hardware co-designed object-oriented VR rendering framework for the parallel rendering in NUMA based multi-GPU system.  

\textbf{Performance Improvement For Rendering.} In order to balance the rendering workloads among multiple GPUs, some studies \cite{moloney2007scalable,hui2009dynamic,eilemann2012parallel} propose a software-level solution that employs CPU to predict the execution time before rendering to adaptively determine the workload size. However, such software-level method requires a long time to acquire the hardware runtime information from GPUs which causes performance overhead. Our hardware-level scheduler could quickly collect the runtime information and conduct the real-time object distribution to balance the workloads. There are also some works \cite{makhinya2010fast, wang2013numa} designing NUMA aware algorithms for fast image composition. Instead of implementing a composition kernel in software, we resort to a hardware-level solution that leverages hardware components to distribute the composition tasks across the multi-GPU system to enhance the pixel throughput. 
Meanwhile, many architecture approaches \cite{arnau2014eliminating,PIMrendering,Perception3D} have been proposed to reduce the memory traffic during rendering. Our work focuses on multi-view VR rendering in multi-GPU system which is orthogonal to these architecture technologies.

\section{Conclusions}

In modern NUMA-based multi-GPU system, the low bandwidth of inter-GPM links significantly limits the performance due to the intensive remote data accesses during the multi-view VR rendering. In this paper, we propose object-oriented VR rendering framework (OO-VR) that converts the remote inter-GPM memory accesses to local memory accesses by exploiting the data locality among objects. First, we characterize the impact of several parallel rendering frameworks on performance and memory traffic in the NUMA-based multi-GPU systems. We observe the high data sharing among some rendering objects but the state-of-the-art rendering framework and multi-GPU system cannot capture this interesting feature to reduce the inter-GPM traffic.
Then, we propose an object-oriented VR programming model to combine the two views of the same object and group objects into large batches based on their texture sharing levels. Finally, we design a object aware runtime batch distribution engine and distributed hardware composition unit to achieve the balanced workloads among GPUs and further improve the performance of VR rendering. We evaluate the proposed design using VR featured simulator. The results show that OO-VR improves the overall performance by 1.58x on average and saves inter-GPM memory traffic by 76\% over the baseline case. In addition, our sensitivity study proves that OO-VR can potentially benefit the future larger multi-GPU scenario with ever increasing asymmetric bandwidth between local and remote memory.  

\section*{Acknowledgment}

\vspace{-0.15cm}

This research is supported by U.S. DOE Office of Science, Office of Advanced Scientific Computing Research, under the CENATE project (award No. 466150), The Pacific Northwest National Laboratory is operated by Battelle for the U.S. Department of Energy under contract DEAC05-76RL01830.  
This research is partially supported by National Science Foundation grants CCF-1619243, CCF-1537085(CAREER), CCF-1537062.

%%%%%%% -- PAPER CONTENT ENDS -- %%%%%%%%

%%%%%%%%% -- BIB STYLE AND FILE -- %%%%%%%%
\bibliographystyle{ACM-Reference-Format}
\bibliography{ref}
%%%%%%%%%%%%%%%%%%%%%%%%%%%%%%%%%%%%

\end{document}